\documentclass[%
 reprint,
 superscriptaddress,
 amsmath,amssymb,
 aps,
 pre,
 noeprint,
]{revtex4-2}

\usepackage[caption=false]{subfig}
\usepackage{graphicx}

\DeclareMathOperator{\trunc}{Trunc}

\begin{document}

\title{Modified Axelrod Model Showing Opinion Convergence And Polarization In Scale-Free Networks}

\author{X. Zou}
 \email{xiang.zou@mail.utoronto.ca}
 \affiliation{Department of Physics, The University of Hong Kong, Pokfulam Road, Hong Kong}
 \affiliation{Department of Physics, University of Toronto, Toronto, ON, Canada, }
\author{H. F. Chau}
 \email{Corresponding author, hfchau@hku.hk}
 \affiliation{Department of Physics, The University of Hong Kong, Pokfulam Road, Hong Kong}

\date{\today}

\begin{abstract}
Axelrod model is an opinion dynamics model such that each agent on a square lattice has a finite number of possible nominal opinions on a finite number of issues that are usually called features in the field.  Moreover, its dynamics between two agents is assimilative in the sense that the number of agreeing features between them never decreases upon interaction.  Here we modify this model to study opinion convergence, polarization and more importantly to find ways to reduce opinion polarization in an already polarized population.  We do so by changing or adding several elements from complex network and continuous opinion dynamics research.
 First, we put agents in a scale-free network.  Second, we adopt the bounded confidence model by representing our agent's opinions by numbers in $[-1,1]$ those distances follow the standard Euclidean metric.  Third, our rules allow both convergence and divergence of their resultant opinions after a pair of agents interacts.  As a result, our modified model offers a more comprehensive exploration of opinion dynamics.
 Computer simulation results of our model show scaling behavior and a notable trend in opinion polarization on all features in the majority of reasonable simulation parameters.  To mitigate this polarization, we introduce empathetic agents that work actively to reduce opinion differences. However, our findings indicate limited success in the approach for the most effective way is to change the behavior of a significant portion of highly connected agents.  This research contributes to the understanding of opinion dynamics within society and highlights the nuanced complexities that arise when considering factors such as network structure and continuous opinion values. Our results prompt further exploration and open avenues for future investigations into effective methods of reducing opinion polarization.
\end{abstract}


\maketitle 

\section{Introduction}
\label{introduction}

In socio-physics, opinion dynamics of individuals or groups is modeled via interactions among them. By investigating the emergence and convergence of individual opinions into collective opinions, the socio-physical approach aims to shed light on the underlying mechanisms. In particular, sociophysicists wanted to understand the cause as well as the evolution of opinion formation and dissemination through social influence and social networks.  They also investigate and analyze the conditions that lead to consensus or polarization, and the factors governing the resulting distributions of opinions~\cite{Sociophysics_review}.

Numerous models of opinion dynamics have been proposed throughout the years.  They include the Sznajd model~\cite{Sznajd_model1, Sznajd_model2}, Hegselmann-Krause (H-K) model~\cite{Hegselmann_Krause_model}, Galam models~\cite{Galam_models}, Axelrod model~\cite{Axelrod1997}, Deffuant-Weisbuch model~\cite{Deffuant2000, Weisbuch2002} and their variants~\cite{Sociophysicsmodel1, Sociophysicsmodel2, Sociophysicsmodel3, Sociophysicsmodel4, Chau2014, Sociophysicsmodel1}. Our research here focuses on the Axelrod model. In this model, individuals that we call agents from now on are located on a square lattice. The opinion of each agent, which takes on a discrete set of nominal values, evolves by interacting with neighboring agents according to the set of rules to be described in Sec.~\ref{Sec:Axelrod_Model} below. In particular, Axelrod found that after a sufficiently long time, all agents on the lattice tend to share the same traits for all features provided that the number of features is sufficiently large or the set of all possible opinions is sufficiently small.  Otherwise, multiple regions of different homogeneous opinions on most features is observed in the long run~\cite{Axelrod1997}.  Castellano \emph{et al.} found that by varying the size of the discrete possible opinion set, Axelrod model shows the classical order-disorder phase transition on culture fragmentation when each agent has more than two features~\cite{CMV00}.   Besides, Klemm \emph{et al.} showed that in the thermodynamic limit, adding an arbitrarily small noise rate to the Axelrod model leads to opinion fragmentation~\cite{KETM03}.  Opinions in this model are nominal numbers and its evolution rules on a pair of agents are assimilative in the sense that the number of differing opinions between them does not increase immediately after they interact.  Thus, this model fails to capture the increasing divided and polarized political as well as ideological stances in numerous countries and societies in our contemporary world though modeling realistic opinions was not the original intention of the Axelrod model.

Various groups of social scientists have proposed reasons for this fragmentation and polarization in the real world~\cite{cultural_fragmentation1, political_diverse1, political_diverse2, political_diverse3, Devauchelle2024}. For example, Ashraf and Galor argued that genetic diversity, shaped by the migration of humans out of Africa, is a fundamental determinant of cultural heterogeneity~\cite{cultural_fragmentation1}.  Dandekar \emph{et al.} tried to explain the polarization by opinion formation~\cite{political_diverse1}.

Socio-physical models could offer a complementary approach to understanding the dynamics of opinion fragmentation. Different scholars have tried different simulation models attempting to capture the fragmentation behaviors in society. See, for example, the review by S\^{i}rbu \emph{et al.}~\cite{Sociophysicsmodel1}.  Could the Axelrod model be modified to give fragmented and polarized opinions or to incorporate cultural diversity in the long time limit?  This is the first question we are going to address.  Surely, various authors have studied this issue over the years with partial success. For example, MacCarron \emph{et al.} considered a more complicated interaction rule and analyzed the clustering behaviors. They showed that by decreasing a certain tunable parameter of the model known as the agreement threshold, agents on the lattice will no longer be homogeneous in the sense that they have more than one set of features in the long run~\cite{spreading_culture}.  Radilo-D\'{i}az \emph{et al.} found that for an initial state with small cultural variability, introducing a small amount of rejection dynamics to the Axelrod model is already sufficient to drive the system eventually to a heterogeneous state~\cite{RD09}.  Along another line, Battiston \emph{et al.} proposed that the interactions in real society are carried out in multiple layers, which correspond to different interests or topics.  By considering this multiplicity in their model, they qualitatively presented new dynamical behavior that leads to the emergence and stability of cultural diversity final state~\cite{Battiston2017}.
T\"{o}rnberg proposed a variation of the Axelrod model in which agents have an additional static opinion denoting, say, their partisan or cultural affiliation.  Moreover, they can interact nonlocally.  As expected, dynamic opinions of agents with the same additional static opinion converge.  Interestingly, dynamic opinions of agents with different additional static opinion can be very different.  This so-called sorting behavior appears in spite of the absence of repulsive dynamics~\cite{T22}.
Worth noticing that MacCarron \emph{et al.}~\cite{spreading_culture}, Battiston \emph{et al.}~\cite{Battiston2017} as well as T\"{o}rnberg~\cite{T22} considered discrete feature traits. In contrast, Amorim~\cite{Amorim2014} modified the Axelrod model by aligning it with Bourdieu's theory on the relation between an individual's taste and the social environment.

This paper introduces a socio-physical model of opinion dynamics that builds upon the Axelrod model, incorporating continuous attitude change among agents within a scale-free network using idea borrowed from the bounded confidence model.  First, agents are placed in a more realistic scale-free network of social interaction known as the Barabási–Albert (B-A) network~\cite{BAnetwork}.  Second, instead of using a discrete opinion set, we adopt the bounded confidence model~\cite{Deffuant2000, Hegselmann_Krause_model, Weisbuch2002, JAmodel} by allowing agent's opinions to take on (continuous) values in the closed interval $[-1,1]$ under the standard Euclidean distance metric.  Third, interaction between a pair of agent may lead to either opinion convergence or divergence.  As far as we know, this is the first opinion evolution model that incorporates both assimilative and repulsive dynamics into the Axelrod model with a bounded confidence favor.

Certainly, opinion dynamics studies on scale-free network are not new.  For example, Bartolozzi \emph{et al.} discovered intermittence in a probabilistic two-state model of opinion evolution on the B-A network~\cite{BLT05}.  Jacobmeier studied evolution of discrete cardinal opinion evolution for agents in the B-A network using assimilative dynamics~\cite{Jacobmeier05}.  More importantly, Dinkelberg \emph{et al.} found that the long term behavior of a modified Axelrod model with discrete cardinal opinions is insensitive to network topology~\cite{DMMQ21}.  Similarly, various groups have studied opinion dynamics using the bounded confidence model~\cite{Deffuant2000, Hegselmann_Krause_model, Weisbuch2002, JAmodel}.  (See, for example, the reviews in Refs.~\cite{Lorenz07} and \cite{BCReview} for more details.)  In particular,
Deffuant \emph{et al.} put forward several agent-based models with continuous opinion change such that the involved agents have limited ranges of acceptance and non-commitment, where only the impact of assimilation is taken into account~\cite{Deffuant2000,Weisbuch2002}.  Fortunato \emph{et al.} showed that the eventual states of the combination of the H-K model and the bounded confidence model with assimilative interaction only may consist of multiple homogeneous opinion clustering regions~\cite{FLPR05}.  Lorenz investigated the opinion dynamics of agents placed on $d$-dimensional cubes or simplexes using bounded confidence model with assimilative interaction~\cite{Lorenz08}.  Along a different line, Huet and Deffuant proposed a bounded confidence model with major and minor feature traits whose interaction can be assimilative or repulsive~\cite{HD10}.
Moreover, Chau \emph{et al.}~\cite{Chau2014} proposed a model for opinion formation which is an extension of the Jager-Amblard (J-A) model~\cite{JAmodel} under the framework of bounded confidence.  Chau \emph{et al.}'s model incorporates assimilation and contrast, which influence agents' opinion acceptance or rejection.  More importantly, it shows opinion clustering around extremes and/or moderate. Furthermore, with slow variation of model parameters, Chau \emph{et al.} observed first-order phase transition in the distribution of opinion~\cite{Chau2014}.  Recently, Liu \emph{et al.} researched on the complicated dynamics of the bounded confidence opinion evolution with assimilative and repulsive interactions for agents arranged on a ring~\cite{LMXF23}.
Last but not least, we remark that the works of Jacobmeier~\cite{Jacobmeier05}, Lorenz~\cite{Lorenz08} and Dinkelberg \emph{et al.}~\cite{DMMQ21} mentioned earlier mixed the Axelrod model with the bounded confidence model.  But unlike ours, they did not include repulsive dynamics.

Through computer simulation, we show that agents in our model gradually develop extreme opinions. These findings are consistent with observed social behavior and hence give novel insights into the factors influencing collective human opinions.  Scaling is also observed in the opinion cluster size distribution whose precise definition will be given later in the text.

Most importantly, we also attempted to address our second question, namely, the polarization problem that is so urgently needed in our currently highly divided and politicized society by modifying our model in this paper.
The intuition is that higher openness to ideas from the opposite camp may lead to mutually understanding and increase their willingness to seek common grounds.  Therefore, we set a small portion of agents that are empathetic by actively trying to reduce opinion differences.  (Note that introducing heterogeneous agents has been done before~\cite{Weisbuch2002}, but setting empathetic agents in this way is new.)  Nonetheless, this approach can only economically reduce opinion polarization of an already polarized population by making a significant portion of those highly connected agents empathetic.
This work is an extension of a short project done by XZ~\cite{Myproject} under the supervision of HFC, the teacher of the course ``Data Analysis And Modeling In Physics''.

\section{The Axelrod Model}
\label{Sec:Axelrod_Model}

One of the most frequently used ways of measuring individual opinions in a society is to collect data from attitudinal surveys over a long period of time.  It is instructive to model the attitude evolution of individuals and interest groups.
\label{Axelrod Model}
One such model was proposed by Axelrod~\cite{Axelrod1997}.
In the Axelrod model, agents are placed on a $N\times N$ square lattice with closed boundary conditions.  An agent is conveniently labeled by its position $(i,j)$ on the lattice.
Each agent is characterized by the same fixed set of $F$ features $\mathcal{F} = \{ f_1, f_2, \cdots, f_F \}$.
Each feature takes only values from a fixed set of traits.
For simplicity, each $f_i \in \mathcal{F}$ is taken from the same set of $q$ traits labeled $\{ 0, 1, \dots, q - 1 \}$.
As a result, each agent has $F^q$ possible internal states.
For example, one of the features could be the primary language spoken by the agent, and the corresponding set of traits is the set of all contemporary languages.  Another possible feature is one's religious belief.
Each feature of an agent is randomly assigned in the beginning.
Furthermore, the system evolves for several time steps as follows.
\begin{enumerate}
 \item In each time step, an agent $(i,j)$ is randomly picked.
 \item One of the nearest neighbors of $(i,j)$ is randomly chosen.
 \item Denote the number of features that this pair of agents have in common by $F_\text{comm}$.  And denote the features that the two agents are different by $\mathcal{F}_\text{diff}$.
 \item \label{Rule:change} With probability $P(\text{interaction}) = F_\text{comm}/F$, we randomly pick a feature in $\mathcal{F}_\text{diff}$ and the agent $(i, j)$ then copies the trait of that feature from its chosen neighbor.  In what follows, we call $F_\text{comm}/F$ the cultural similarity between two agents.
\end{enumerate}

Under the Axelrod model, agents' opinions converge provided that $F$ is sufficiently small or $q$ is sufficiently large.

\section{Deficiencies Of The Axelrod Model And Our Remedy}

The Axelrod model is simple and useful in opinion dynamics.  However, it has several deficiencies even though being realistic was not the original goal of the Axelrod model.  We now point them out and report ways to modify the model so as to produce an increasingly polarized opinion distribution that is robust under small changes in tunable parameters.

\subsection{Network Topology}
\label{Subsec:topology}
In the Axelrod model, agents form a two-dimensional square lattice.  This network topology is artificial.  More importantly, it is not compatible with the fact that many social structures can be described by scale-free networks~\cite{scalefreesociety1}.  In simple terms, all but those agents in the lattice boundary in the Axelrod model have equal strength in influencing others or being influenced by others.  In contrast, scale-free network topology can be used to model both social and parasocial interactions in the real world, including interactions between social media influencers and their followers.  Thus, we propose putting agents in an (undirected) scale-free network in which only neighboring agents are allowed to interact directly.  For simplicity, we generate such a network using the B-A model~\cite{BAnetwork} as it is one of the most studied and used scale-free networks.  Besides, it has the additional advantage that it can be efficiently generated.  In this study, we simplify matters by considering non-evolving networks.  It is instructive to consider evolving networks in future studies.

\subsection{Feature Traits}
\label{Subsec:opinion}
In the Axelrod model, every agent is described by a set of features, and each feature can have several possible discrete nominal traits~\cite{Axelrod1997}.   While there are features that are naturally described by discrete nominal traits, the Axelrod model is incomplete as traits of a feature need not be discrete or nominal.  For example, one of the features could be the agent's fluency in a particular spoken language.  Here, the possible traits are more naturally described by a continuous real number than discrete nominal states.  More importantly, using continuous traits is a natural platform to model the possibly increasingly polarized opinion distribution.  We, therefore, modify the model by replacing the discrete set of traits with a real number representing the strength of the agent's feature.  We also call this type of continuous feature traits opinions of the agent.

How could we take care of the increasingly polarized opinion distribution?  A simple and effective way to address this issue is to borrow the idea of the bounded confidence model~\cite{Deffuant2000, Hegselmann_Krause_model, Weisbuch2002, JAmodel} by representing opinions using a real number in a bounded interval under the Euclidean metric.  For simplicity, we set the possible opinions to be a real number in the interval $[-1,1]$.  Besides, we denote the distance between two opinions values $p$ and $q$ by
\begin{equation}
 d(p,q) = |p-q| .
 \label{E:d_Euclidean}
\end{equation}

As for the initial values of each trait, instead of drawing from a uniform distribution in $[-1,1]$, we use a more realistic truncated normal distribution with respect to the Euclidean distance metric.  Its detailed construction can be found in upcoming subsection.

\subsection{Interaction Rules Among Agents}
\label{Subsec:interaction}
Recall that agents change their traits according to rule~\ref{Rule:change} in the Axelrod model. 
This rule is assimilative and hence can only increase the cultural similarity between the chosen pair of agents.  In other words, rule~\ref{Rule:change} is too cohesive.  Besides, it is not consistent with the observation that cultural diversity among groups of people appears even in apparently culturally homogeneous societies~\cite{cultural_fragmentation1, political_diverse1, political_diverse2, political_diverse3}.

We may also need to take repulsive dynamics into consideration.  The existence of repulsive dynamics and its psychological origin (if it exits) are current research topics.  One possibility is that the effect is real.  Supporters of this camp believe that social interaction where attempts to persuade or restrict someone's freedom of opinion could result in the adoption of an opposing view. This effect occurs when persuasive communication happens in an unintended position, which often strengthens the original attitude an individual holds that it sought to alter.  Hovland \emph{et al.} first identified and named it the boomerang effect~\cite{HJK53}.
The boomerang effect has been observed in various contexts \cite{Boomerang_effect1, Boomerang_effect2}, including consumer behavior, psychological interventions, health policies, and, relevant to our discussion here, political communication. In political contexts, attempts to correct misinformation through mock news articles sometimes lead to increased belief in inaccurate information among certain groups~\cite{Boomerang_misinformation_increase}.  Nevertheless, we need to the caution here as the evidence on the existence of the boomerang effect is controversial and mixed.  Furthermore, this effect may not be systematic.
For example, Schultz \emph{et al.} observed it in a field experiment in promoting household energy conservation through the use of different normative messages.   Moreover, boomerang effect was eliminated by adding an injunctive message~\cite{SNCGG07}.  Similar conclusions were drawn in the follow up study by Verkooijen \emph{et al.}~\cite{TSM15}.  In addition, Liu \emph{et al.} found that overinvestment in corporate social responsibility had a boomerang effect on the shareholder value of a company after announcing a recall.  The effect is more evident when institutional ownership is low or when customer awareness is high~\cite{LLWX19}.
On the other hand, Tak\'{a}cs \emph{et al.} failed to observe it in their recent study using the University of Groningen students as participants~\cite{TFM16}.
Bail \emph{et al.} discovered that exposing Democrats (Republicans) with conservative (liberal) social media messages make their views more liberal (conservative).  However, the change is statistically insignificant~\cite{Bail18}.
Another possibility is that boomerang effect is neither real nor systematic.  At least, agents tend to be less affected by ideas from the opposite camp~\cite{T22}.  If it is so, what is the underlying mechanism for opinion polarization?  Unfortunately, no definitive answer has been identified.

Here we move on by adopting an empirical approach.  That is to say, we model the repulsive dynamics as if it were coming from the boomerang effect without considering its psychological origin.  Our goal is to reproduce the polarization seen in the real world and study if there is anything we can do to reduce it.
Specifically, to study and model the positive persuasion as well as the repulsion effects in social interactions, we introduce a parameter $a \ge 0$ known as the agreement threshold.  It is used to determine whether a social interaction results in opinion convergence or divergence.
In each time step, we randomly pick an agent from the scale-free network plus one of its neighbors.  We count the number of features in which the Euclidean distance between the opinions of the two agents is smaller than the agreement threshold $a$.  If this number is greater than or equal to, say, half of the total number of features, the two agents are said to be ``similar''.  Otherwise, they are ``dissimilar''.
We hypothesize that upon interaction, opinions of similar agents are likely to converge whereas those of dissimilar ones are likely to diverge.  Besides, an agent's persuasion power is proportional to the number of neighbors it has.  This models the asymmetric influence between agents of different connectivities. 
This hypothesis is consistent with a number of findings in the literature on the persuasion power of social media influencers.  They include the meta-analysis by Pan \emph{et al.} in which they found that the characteristics of a social media influencer, including the number of followers, is positively correlated with the purchasing behavior of his/her followers~\cite{PBGL25}.
The report by Conde and Casais showed the positive correlation of the influencing power with the perceived popularity and opinion leadership from a sample of 140~Portuguese social media influencers~\cite{CC23}.  The work by Peter and Muth who discovered that social media influencers were increasingly affecting German youth on political issues.  The effects ranged from amplifying the effects of existing opinions to opinion formation as well as to change voting intentions~\cite{PM23}.  Last but not least, the work by Liu and Zheng who argued that the influence between social media influencers and their followers is mutual when evaluating brand credibility and purchase intention.  They further found that the influence could be repulsive sometimes~\cite{LZ24}.  Nevertheless, we have to be caution that little is known on how the followers directly affecting the positions of social media influencers.  It is instructive and timely to investigate the validity of the hypothesis that an agent's persuasion power is proportional to its number of neighbors in the field experiments.

After the above preparation and discussion, we now state the details of our model below.
\begin{enumerate}
 \item {[Scale-free network generation]} A B-A network with $N$ vertices is generated with each newly added vertex connecting to $m$ existing nodes using their standard preferential attachment mechanism~\cite{BAnetwork}.  Each vertex of the network represents an agent and only those agents joined by an edge in the network can communicate directly with each other.
 \item {[Opinion initialization]} Each agent is characterized by the same finite set of $F$ features $\mathcal{F} = \{ f_1,\cdots,f_F \}$.  The opinion of agent $X$ on feature $f \in \mathcal{F}$, denoted by $X_f$ is randomly and independently drawn from $\mathcal{N}_t(\bar{x},\sigma) = \mathcal{N}_t(0,0.25)$, namely, the truncated normal distribution with mean $0$ and standard deviation $0.25$ in the Euclidean distance metric.  More precisely, if $r$ is the value of the random variable drawn from $\mathcal{N}_t(0,0.25)$, then the opinion is set to $r$ if $|r| \le 1$.  And the opinion is set to $1$ $(-1)$ if $r > 1$ ($r < -1$).
 \item \label{Rule:evolution} {[Opinion evolution]} An agent $X$ is randomly picked.  Then, a neighbor of $X$, say $Y$, is randomly selected.  A feature $f$ is randomly selected from $\mathcal{F}$.
 \begin{enumerate}
  \item {[The case of similar agents]} If the two agents are similar, namely, the Euclidean distances between at least $\kappa = 1/2$ of their opinions are less than the agreement threshold $a$, then their opinions of feature $f$ move closer to each other in the sense that the Euclidean distance of the new opinions of the selected feature trait between the two agents decreases upon interaction.  Specifically, $X_f$, the opinion of feature $f$ for agent $X$, changes to $\tilde{X}_f$ where
   \begin{subequations}
    \label{E:opinion_evolution}
    \begin{equation}
     \tilde{X}_f = X_f - \frac{\mu c_Y |X_f - Y_f|}{c_X + c_Y} .
     \label{E:opinion_evolution_X}
    \end{equation}
    Here, $c_X$ and $c_Y$ are the degrees of nodes $X$ and $Y$, respectively.  In addition, $\mu > 0$ is a parameter modeling the effectiveness of persuasion.  Note that we borrow the preferential attachment idea in the B-A network~\cite{BAnetwork} in this rule so as to model the effective persuasion power of social media influencers.  Likewise, $Y_f$ changes to $\tilde{Y}_f$ where
    \begin{equation}
     \tilde{Y}_f = Y_f - \frac{\mu c_X |Y_f - X_f|}{c_X + c_Y}
     \label{E:opinion_evolution_Y}
    \end{equation}
   \end{subequations}
  \item {[The case of dissimilar agents]}  If the two agents are dissimilar (that is, they are not similar), then we randomly pick a feature $f$ among those with $d(X_f,Y_f) \ge a$.
   The values of $X_f$ and $Y_f$ change to $\tilde{X}_f$ and $\tilde{Y}_f$ according to
   \begin{subequations}
    \begin{equation}
     \tilde{X}_f = \trunc \left[ X_f + \frac{\mu c_Y |X_f - Y_f|}{c_X + c_Y} \right]
    \end{equation}
    and
    \begin{equation}
     \tilde{Y}_f = \trunc \left[ Y_f + \frac{\mu c_X |Y_f - X_f|}{c_X + c_Y} \right] ,
    \end{equation}
   \end{subequations}
   respectively.  Here
   \begin{equation}
    \trunc (p) =
    \begin{cases}
     p & \text{if~} |p| \le 1 , \\
     1 & \text{if~} p > 1, \\
     -1 & \text{if~} p < -1 .
    \end{cases}
   \end{equation}
   Clearly, the Euclidean distance of the new opinions of the selected feature trait between the two agents increases or remains unchanged upon interaction.
 \end{enumerate}

  The opinion evolution in Step~\ref{Rule:evolution} is repeated $N t/2$ times, where $t$ is the average number of interactions an agent has with its neighbors.
\end{enumerate}

Note that we use the probability distribution $\mathcal{N}_t(0,0.5)$ to initialize opinions to avoid opinion bias as well as to give an initial opinion diversity in our initial ensemble.
In addition, our evolution rules do not change the nature of the agent pairs picked in Step~\ref{Rule:evolution} in the sense that they are similar (dissimilar) before and after the interaction.  Nevertheless, the nature of each of the picked agents with its other neighbors may change after the interaction.
Last but not least, we stress that for all the subsequent simulations that study the eventual system behavior, the average number of interactions an agent has with its neighbors $t$ is chosen to be large enough for the system to approach equilibration.

\section{Simulation Results And Effects Of Different Model Parameters}
\label{Sec:simulation}
\subsection{Methodology}
\label{Subsec:methodology}

We study the dynamics of our modified Axelrod model using Python~\cite{pythoncode}.  The scale-free network is created using the Networkx package and most of the analytical work is done using the NumPy package.  Figures are produced by the Matplotlib package.

\begin{table*}[!t]
    \centering
    \begin{tabular}{|p{2cm}|p{9.5cm}|p{2cm}|}
    \hline
    Parameter & Description & Default value \\
    \hline
    $N$ & Number of agents & 1000\\
    \hline
    $m$ & Number of edges attached from new node to existing nodes while constructing the initial B-A network & 4 \\
    \hline
    $F$ & Number of features & 11\\
    \hline
    $t$ & Mean number of interactions per agent & 200 \\
    \hline
    $\mathcal{N}_t(\bar{x},\sigma)^{\vphantom{M^2}}$&
    Initial probability distribution of opinions
    & $\mathcal{N}_t(0,0.25)$\\
    \hline
    $a$ & Agreement threshold that determines the similarity of agents & 0.25\\
    \hline
    $\kappa$ & Fraction of feature traits used to determine the similarity of agents & 0.5\\
    \hline
    $\mu$ & Persuasion effectiveness parameter & 0.25 \\
    \hline
    \end{tabular}
    \caption{Default values of different parameters used in our simulation.}
    \label{T:parameters}
\end{table*}

\begin{figure*}[!t]
     \centering
     \subfloat{
         \includegraphics[width=\textwidth]{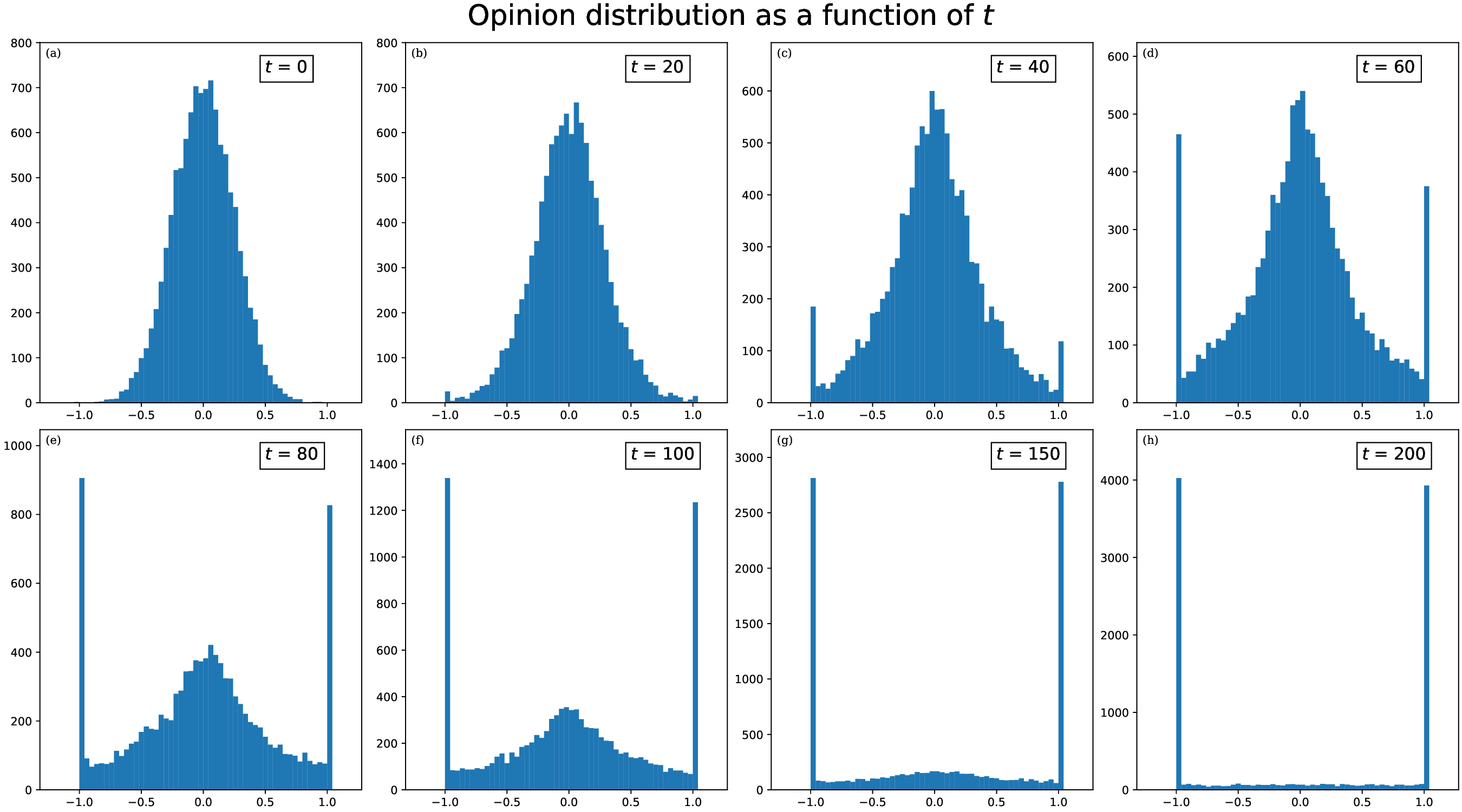}
	}
	\caption{Histograms showing the time evolution of opinion distribution with all but the parameter $t$ taken from Table~\ref{T:parameters}.  The $x$-axes of all subplots are the opinions in $[-1,1]$ and the $y$-axes are the occurrence frequencies of the opinion. Although the initial opinions concentrate around $0$, the system evolves to a state of extreme opinion polarization in the long time limit in which nearly all agents have extreme opinions on all features.  Note that the $y$-axis ranges of all histogram subplots in this paper may be different.}
        \label{fig:normal_vary_steps}
\end{figure*}

We investigate the effect of each model parameter on the development of the system by varying one of them and fixing all the rest to their corresponding default values as tabulated in Table~\ref{T:parameters}. 
Each of the simulation results presented below is taken from a single run randomly selected over a sample of 10~runs.  And a new scale-free network is used for each run.  In fact, all runs give very similar distributions.
We report our simulation results in histograms on the opinion distribution of feature traits.  Since our initialization and evolution rules apply to different feature traits in the same way, opinion distributions of different feature traits averaged over a sufficiently large ensemble of independent runs should be the same.  This is indeed what we have observed in our simulation.  Consequently, to save running time, we lump the distributions of all $F$ feature traits of every agent in plotting the histogram.  In each histogram, we divide the opinion into $50$~evenly spaced bins (in the usual Euclidean distance) each with a bin width of~$0.04$.

\subsection{Feature Evolution}
We begin by investigating the time evolution of opinions by fixing all other model parameters to the ones listed in Table~\ref{T:parameters}.
Fig.~\ref{fig:normal_vary_steps} shows the time evolution of agents opinions. Clearly, opinion spreads as time progresses.  When the mean number of interactions per agent $t$ is less than $\approx 20$, the opinion distribution roughly follows the Gaussian distribution whose mean is zero and the standard deviation increases gradually. When $t \gtrsim 20$, extreme opinions start to emerge.  In addition, the fractional increase of extreme opinions accelerates.  Eventually, extreme opinions dominate, and nearly all agents have extreme opinions on all features.  That is to say, opinions become extremely polarized.

\begin{figure*}[!t]
     \centering
     \subfloat{
         \includegraphics[width=\textwidth]{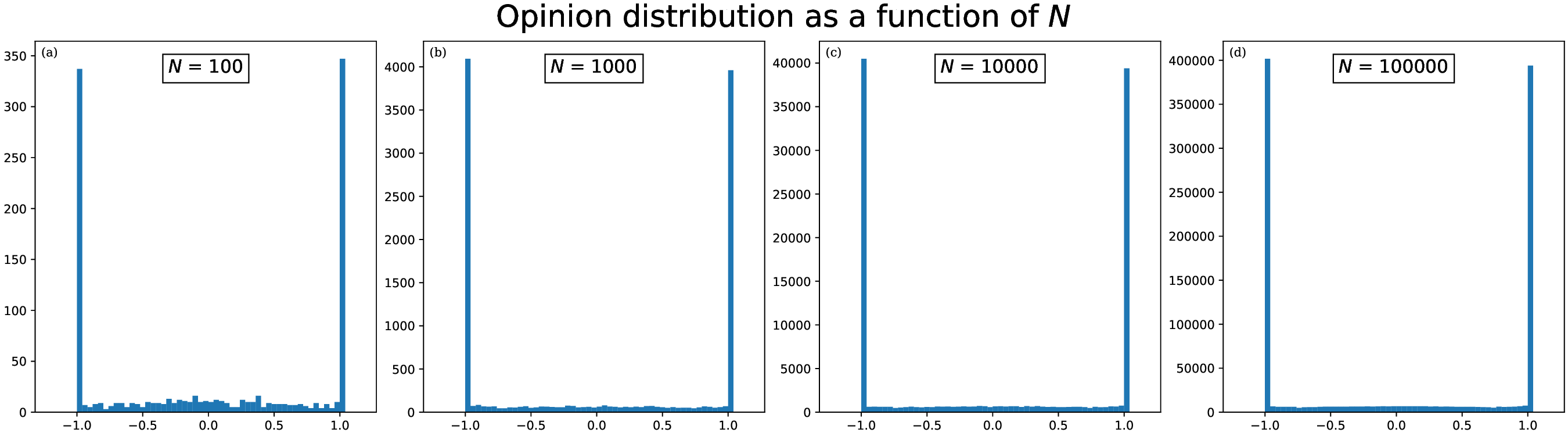}
	}
	\caption{Opinion distributions after $N t/2$ time steps with $t = 200$ for different values of $N$.  These subplots show no significant difference in opinion distribution for different $N$.}
        \label{fig:normal_vary_agentsno}
\end{figure*}

Our simulation result can be qualitatively understood as follows.  Since the agreement threshold $a$ is less than the standard deviation of the initial opinion $\sigma$, a sizable fraction of neighboring agents is dissimilar.  Their interactions gradually lead to opinion polarization.  Once a non-negligible portion of the opinion is polarized, our evolution rules accelerate further polarization for it is harder and harder to find a similar agent pair which consists of an agent with moderate opinion and another agent with an extreme opinion on the same feature.  As a consequence, it is no longer possible to pull enough extreme opinion agents back to moderate ones.  Thus, an extremely polarized society is formed in the end.
Readers will find in our subsequent analysis that the formation of extremely polarized opinion is a genuine and common feature of our model using reasonable parameters rather than the result of a cherry-picked set of parameters.  (Surely, opinions are never polarized if $\sigma \approx 0$ or $\kappa \approx 0$.  However, these sets of parameters are atypical and unrealistic.)

 Let us compare our opinion distribution results with several models on market.  The three closest models on market are the ones reported by Chau \emph{et al.}~\cite{Chau2014}, Jager and Amblard~\cite{JAmodel} as well as Liu \emph{et al.}~\cite{LMXF23}.  They used Euclidean metric bounded confidence with assimilation as well as contrast dynamics.  Agents are placed on a complete graph in Refs.~\cite{Chau2014,JAmodel}, regular 2-d lattice in Ref.~\cite{JAmodel} and a ring in Ref.~\cite{LMXF23}.  Using suitable parameters, the eventual agent opinion distributions in these two models exhibit bipolar extreme polarization very similar to ours.

\subsection{Effect Of The Number Of Agents}
\label{Subsec:effect_N}
We now study the effect of the number of agents $N$ while fixing other parameters to their default values listed in Table~\ref{T:parameters}.  As shown in Fig.~\ref{fig:normal_vary_agentsno}, there is no significant difference in the distribution between different histograms even if $N$ varies over four orders of magnitude from $10^2$ to $10^5$. This result is not surprising because, upon coarse-graining, the resultant network is still scale-free.  And since the formation of extreme opinion polarization is the generic behavior of the system, the effective parameters of the renormalized system likely fall within this generic region. Finally, we find that the convergence rate of the opinion polarization is almost independent of $N$.  That is to say, for a given mean number of interactions per agent $t$, the distribution between different histograms is statistically the same regardless of $N$.  

\begin{figure*}[!ht]
 \includegraphics[width=\textwidth]{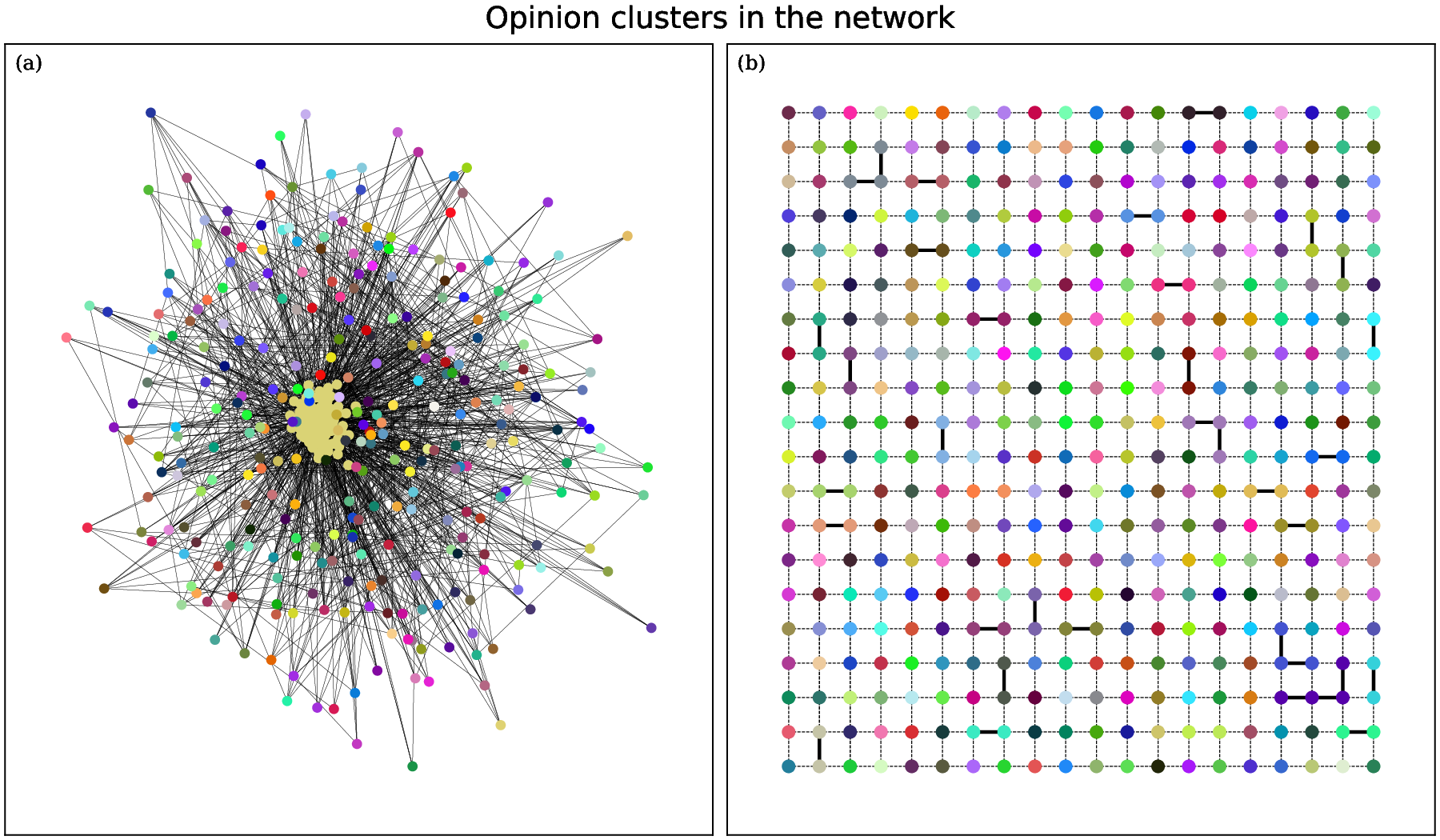}
 \caption{Opinion cluster distributions of a typical run of our model on (a)~a B-A network and (b)~$20\times 20$ lattice using parameters in Table~\ref{T:parameters} except that $N = 400$.  Agents in the same opinion cluster are represented by dots of the same color; and no adjacent cluster is of the same color.  Each adjacent agent pair in~(a) is linked by a solid black line.  In~(b), a thick solid black (thin dashed black) line is drawn between a pair of adjacent similar (dissimilar) agents.}
        \label{fig:opinion_clusters}
\end{figure*}

\subsection{Distributions Of Opinion Clusters And Their Sizes}
\label{Subsec:opinion_cluster}
\begin{figure*}[!ht]
 \includegraphics[width=\textwidth]{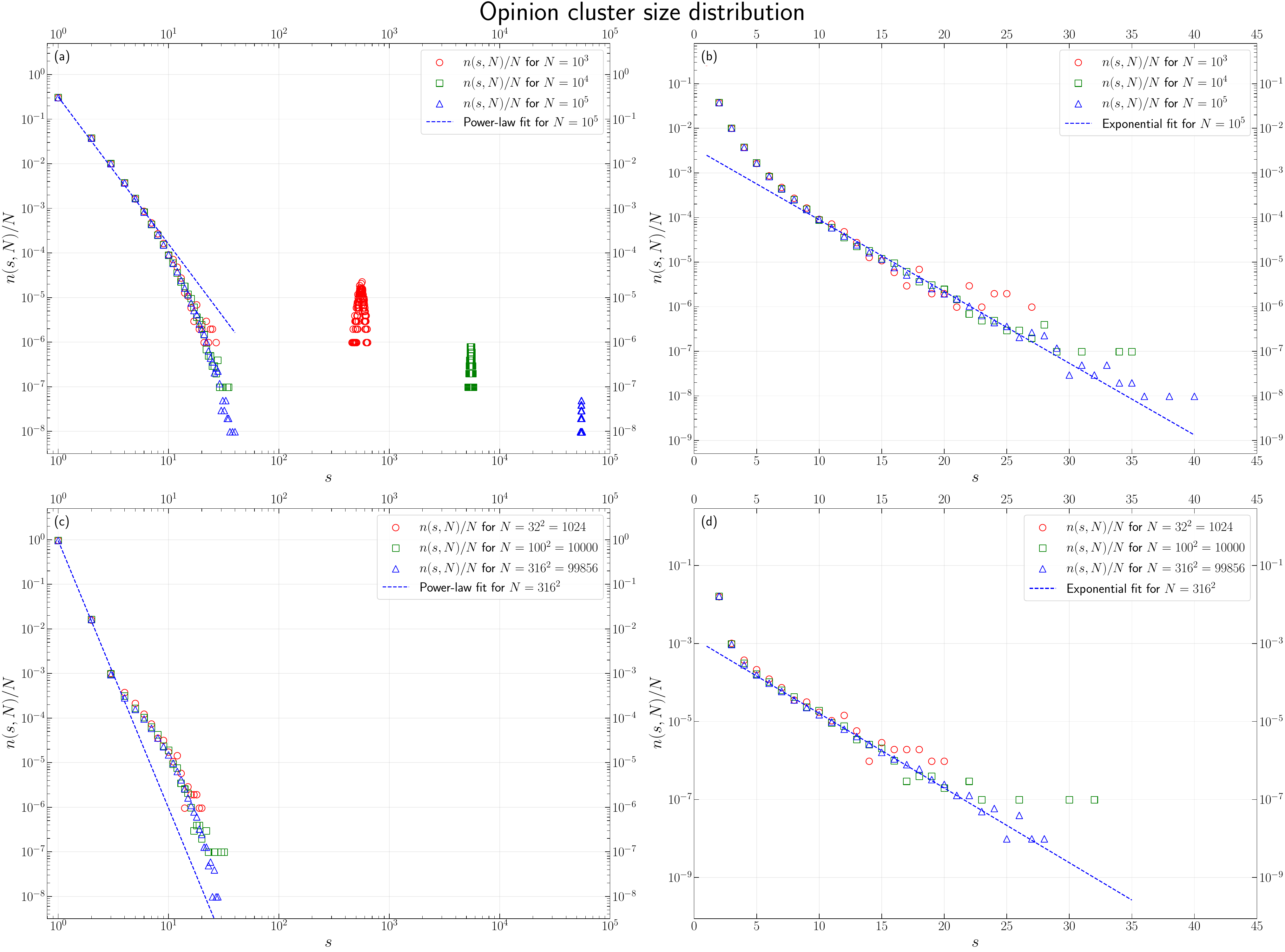}
 \caption{The average number of opinion cluster divided by the number of agents $n/N$ as a function of cluster size $s$ and number of agents $N$ on a B-A network in (a)~log-log and (b)~semi-log plots as well as on a square lattice in (c)~log-log and (d)~semi-log plots using parameters in Table~\ref{T:parameters} except that $N$ can vary.  The red dots, green squares and blue triangles are for $N = 10^3$, $10^4$ and $10^5$, respectively for the B-A network and the responding closest square number for the square lattice.  Each set of data is obtained by averaging over $1024$ independent runs.  The uncertainty of $\log_{10}[n(s,N)/N]$ of each data point is of order of $(N S)^{-1/2}/n$ and the corresponding error bar is omitted in the subfigures for clarity.  The dashed line in each subfigure is the best fit curve whose parameters are shown in Table~\ref{T:fit_parameters}.}
        \label{fig:opinion_clusters_power_law}
\end{figure*}

We turn to study the opinion cluster distribution and its relation with $N$ where an opinion cluster is defined to be a maximal (in the sense of set inclusion) path-connected set of agents such that any pair of adjacent agents (with respect to the underlying network topology) in the set is similar.  Fig.~\ref{fig:opinion_clusters}(a) shows the distribution of opinion clusters for a typical run using parameters in Table~\ref{T:parameters} at $t = 200$ with $N$ replaced by $400$ to avoid over-crowding the graph.  We find that in about $2/3$ of the runs, one dominant opinion cluster consisting of about 50\% to 60\% of the population shows up.  (One example is Fig.~\ref{fig:opinion_clusters}(a) whose dominant opinion cluster is made up of the pale yellow dots in the central region of the figure.)  Whereas for the remaining $1/3$ of the runs, one large and one much smaller dominant opinion clusters appear along side with many small clusters.  Besides, Fig.~\ref{fig:opinion_clusters_power_law}(a) tells us that as $N$ increases, both the chance of having more than one dominant opinion cluster and the standard deviation of the size of the dominant opinion cluster decrease.  For instance, when $N = 10^5$, the distribution of the size of the dominant opinion cluster sharply peaks in the interval $[0.54 N, 0.56 N]$ up to $3\sigma$ level.  Furthermore, for all the number of agents $N$ used in our simulation, every remaining cluster contains much fewer number of agents.  In other words, there is no intermediate sized opinion cluster in the system.  To summarize, a highly fragmented society with a significant number of highly connected agents holding similar extreme opinions is formed.  More importantly, numerous small clusters many of them holding opposite extreme opinions are also present.  This polarization pattern resembles those observed in the real world such as the great opinion differences between those living in cities and in rural areas as well as the emergence of political extremism due to distrust of the elites.  It is instructive to explore the potential of applying our model to study this type of society as the general trend reported above is generic over typical sets of parameters.
In contrast, Fig.~\ref{fig:opinion_clusters}(b) shows the opinion clusters of a typical run of our model when agents are placed in a $20\times 20$ lattice using the same parameters as in Fig.~\ref{fig:opinion_clusters}(a).  Unlike the case of using B-A network, here we see a much greater number of similarly sized opinion clusters.  Besides, no dominant cluster is present.  (For instance, the largest opinion cluster in Fig.~\ref{fig:opinion_clusters}(b) consists of $4$ agents represented by purple dots near the bottom right corner of the figure.)

We further investigate the opinion cluster size distribution.  Let us denote the average number of opinion cluster of size $s$ in a system of $N$ agents by $n(s,N)$.  Figs.~\ref{fig:opinion_clusters_power_law}(a) and~(b) show the log-log and semi-log plots of $n(s,N)/N$.  The data points are obtained for $N = 10^3, 10^4$ and $10^5$ each averaged over at least $S = 1024$ runs using parameters listed in Table~\ref{T:parameters} except that $N$ can vary.  Scaling is observed in $n(s,N)/N$ as curves for different values of $N$ collapse.  Moreover,
\begin{equation}
 \frac{n(s,N)}{N} \approx
 \begin{cases}
  A s^{-\tau} & \text{for~} s \lesssim 10, \\
  B e^{-\lambda s} & \text{for~} 15 \lesssim s \lesssim 35 ,
 \end{cases}
 \label{E:power_law_fit}
\end{equation}
 with fitting parameters $A, \tau, B$ and $\lambda$ tabulated in Table~\ref{T:fit_parameters}.  Figs.~\ref{fig:opinion_clusters_power_law}(a) and~(b) depict that opinion cluster distribution follows a power law over roughly one decade for $s \lesssim 10$ and an exponential decay law over roughly two decades for $15 \lesssim s \lesssim 35$.  The power law distribution for small cluster size can be understood as follows.  Since $n(s,N)$ obeys the sum rule $\sum_s s \ n(s,N) = N$ and agents are placed on a scale-free network, we expect that
 $n(s,N)/N \sim s^{-\tau}$
for sufficiently small $s$.

\begin{table}[t]
    \centering
    \begin{tabular}{|c|c|c|}
    \hline
    Parameter & B-A network & square lattice \\
    \hline
    $\log_{10} A$ & $-0.51\pm 0.02$ & $-0.02\pm 0.06$ \\
    \hline
    $\tau$ & $3.2\pm 0.2$ & $5.9\pm 0.4$ \\
    \hline
    $\log_{10} B$ & $-2.4\pm 0.1$ & $-2.9\pm 0.1$ \\
    \hline
    $\lambda$ & $0.16\pm 0.01$ & $0.19\pm 0.01$ \\
    \hline
    \end{tabular}
     \caption{Best fit parameters for the subplots in
      Fig.~\ref{fig:opinion_clusters_power_law}.}
    \label{T:fit_parameters}
\end{table}

 Next, we explore the effect of the underlying network topology on opinion cluster size distribution.  Figs.~\ref{fig:opinion_clusters_power_law}(c) and~(d) show the log-log and semi-log plots of $n/N$ as a function of $s$ for different $N$ when agents are placed on a square lattice.  Just like the B-A network case, scaling is observed as the curves collapse for any $N$.  But unlike the B-A network situation, power law is observed in a much smaller region of $s \lesssim 4$ --- a region too small to be of statistical significance.  Furthermore, exponential decay is found for $5 \lesssim s \lesssim 25$.  The corresponding best fit parameters are listed in Table~\ref{T:fit_parameters}.  Therefore, it appears that the use of scale-free network enlarges the power law regime of opinion cluster distribution.

We now compare our opinion cluster size distribution results with a few existing models in the literature.  Opinion cluster size distribution of our model in B-A network is similar to that by Jacobmeier~\cite{Jacobmeier05} that also used B-A network but only with assimilative dynamics.  Both shows power law distribution but with very different exponents over very different regions --- their exponent is close to $0$ and is valid for intermediate values of $s$ over a range of about $1.5$~decades while our exponent is about $3.4$ and is valid for $s \lesssim 10$.  In contrast, their model shows clustering in opinions.
Furthermore, our model using B-A network shares similar opinion cluster size distribution with the model by Deffuant \emph{et al.}~\cite{Deffuant2000} that used assimilative dynamics only in a regular 2-d lattice.  However, their distribution exhibits power law but with no obvious scaling.  Our 2-d square lattice result is also similar to the stable configuration of the polarization phase of the J-A model \cite{JAmodel}.  Besides, ours exhibits scaling but without any power law.  In addition, the behavior of our model is different from that in Dinkelberg \emph{et al.}~\cite{DMMQ21} as the cluster size distribution of the later is weakly dependent on the underlying network used.

\begin{figure*}
    \centering
    \includegraphics[width=\textwidth]{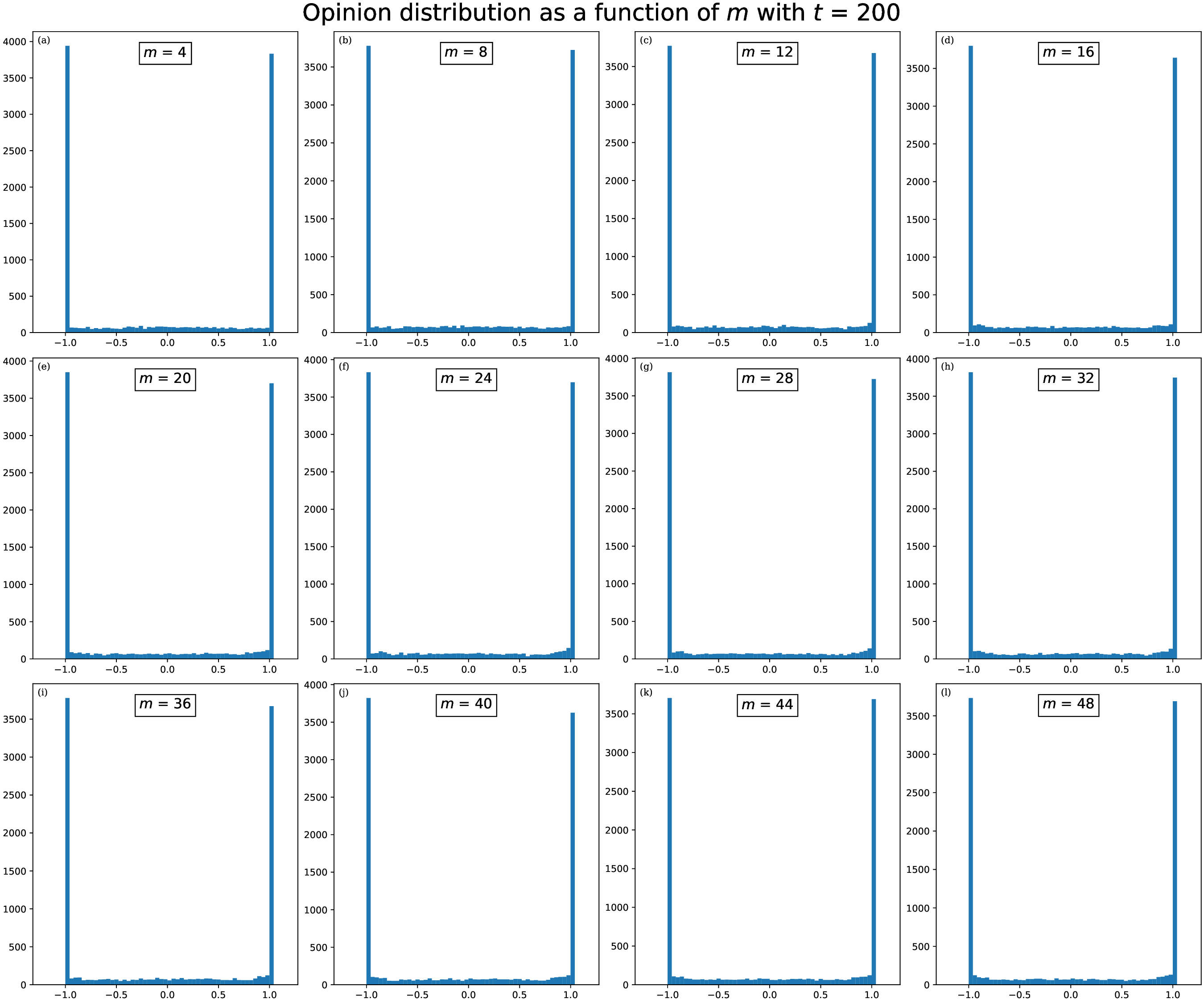}
    \caption{Opinion distribution after $N t /2$ time steps with $N = 1000$ and $t = 200$ for different values of $m$.  These subplots show no significant difference in opinion distribution for different $m$.}
    \label{fig:normal_vary_agentdeg_200_steps}
\end{figure*}

\subsection{Effect Of The Average Degrees Of Agents In The Scale-free Network}
Here we investigate the effect of the average degree of the scale-free network $m$ on the system evolution by varying it from $4$ to $48$.  As shown in Fig.~\ref{fig:normal_vary_agentdeg_200_steps}, there is no significant difference in the final steady state of the opinion distribution among all agents.  They all evolve to a highly polarized opinion state.

\begin{figure*}
    \centering
    \includegraphics[width=\textwidth]{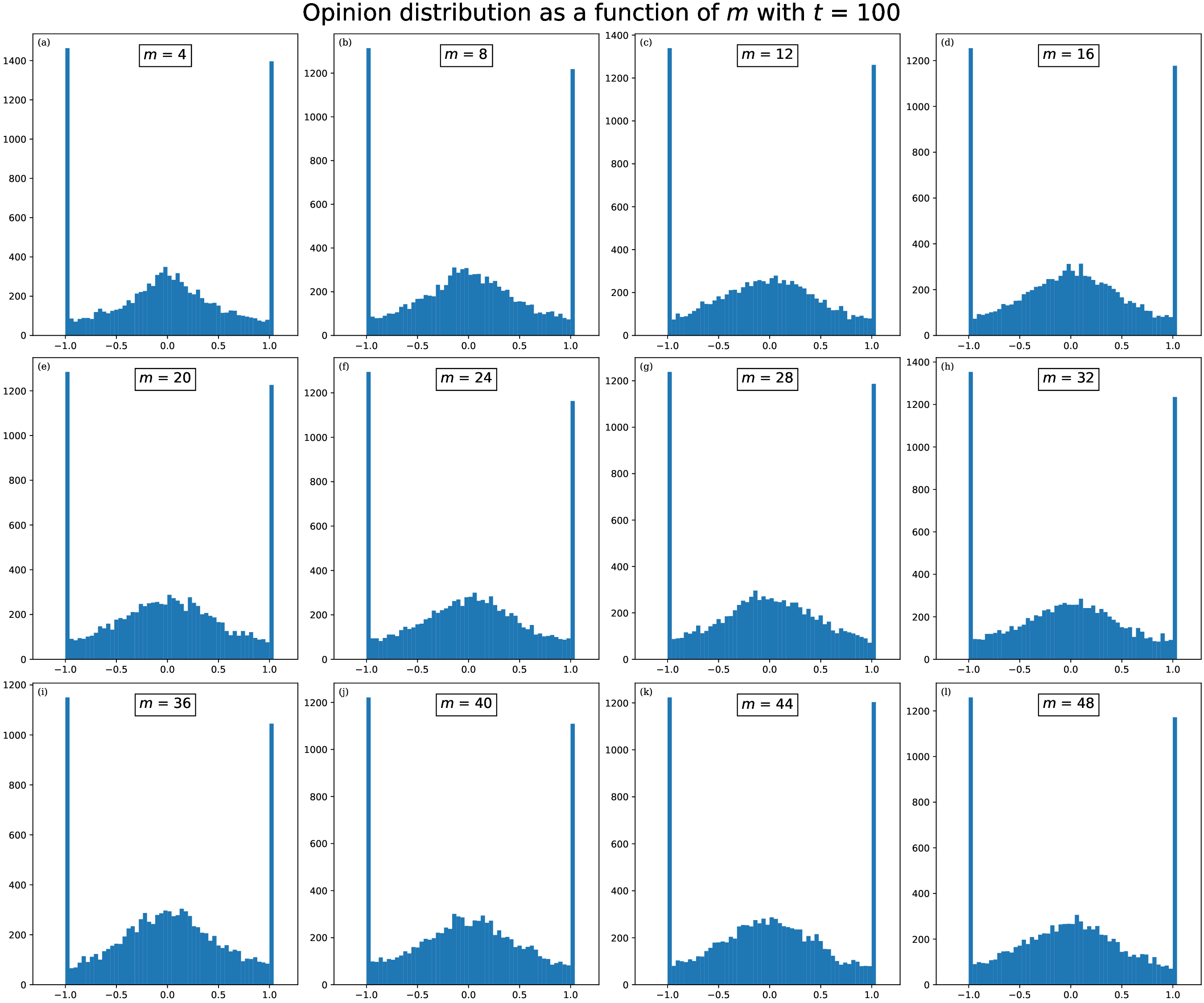}
    \caption{Opinion distribution after $N t/2$ time steps with $N = 1000$ and $t = 100$ for different values of $m$.  Compared with Fig.~\ref{fig:normal_vary_agentdeg_200_steps}, these subplots show that longer equilibration times are needed to reach the extremely opinion depolarization state as $m$ increases.}
    \label{fig:normal_vary_agentdeg_100_steps}
\end{figure*}

To study the rate of convergence to this highly polarized state, we reduce the average number of interactions per agent $t$ from $200$ to $100$.
Fig.~\ref{fig:normal_vary_agentdeg_100_steps} depicts that the equilibration time to the extremely depolarized state increases as $m$ increases.  This result can be understood as follows.  An agent has more neighbors on average as $m$ increases.  Consequently, the opinion of an agent is influenced by more neighbors each has statistically independent opinions on each feature trait initially.  Thus, it takes a longer time for opinions of the system to polarize.

\subsection{Effect Of The Agreement Threshold}
\label{Subsec:effect_a}
\begin{figure*}
    \centering
    \includegraphics[width=\textwidth]{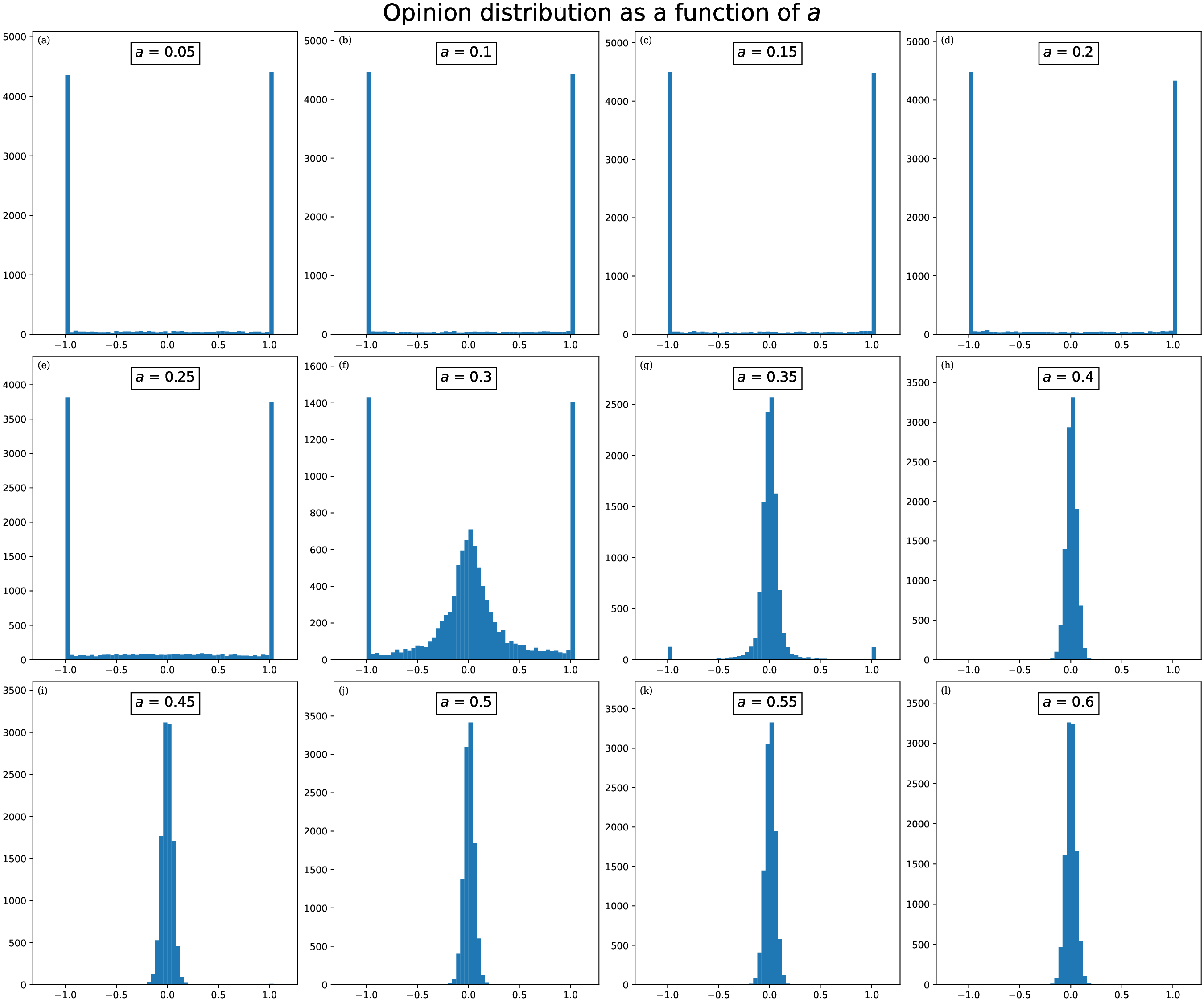}
    \caption{Opinion distribution after $N t/2$ time steps for $N = 1000$ and $t = 200$ for different values of $a$.  It shows that opinion polarization occurs only when $a \lesssim \sigma$.}
    \label{fig:normal_vary_agreement}
\end{figure*}

We now study the effect of agreement threshold $a$ on the system.  When $a \lesssim \sigma$, as shown in subplots~(a)--(e) in Fig.~\ref{fig:normal_vary_agreement}, the system equilibrates to two dominating peaks at a polarized opinion as in the generic case.  This is because the initial opinion distribution is sufficiently diverse to allow a sizable portion of neighboring dissimilar agents.  Their interactions seed more extreme opinions.  In contrast, when $a \gtrsim \sigma$, the system evolves to a single peak of moderate opinion and the extreme opinions population is very small.  (See subplots~(h)--(l) in Fig.~\ref{fig:normal_vary_agreement}.)  The reason is that with a large value of $a$, almost all agent pairs are similar.  This leads to opinion convergence for at least a majority portion of the population.  For $a \approx \sigma$, as depicted in subplots~(f) and~(g) in Fig.~\ref{fig:normal_vary_agreement}, the coexistence of a non-negligible number of extreme and moderate opinions is observed.

In summary, we find that the value of $a/\sigma$ is the most important factor determining the asymptotic behavior of the system.  It is natural to expect that $a < \sigma$ in human societies for otherwise the repulsive dynamics can hardly be observed.  The implication of our modified Axelrod model is that societies will evolve toward extreme opinion polarization --- a phenomenon in our contemporary society that leads to all kinds of troubles making most people unhappy and some to go violence.

\subsection{Effects Of Other Parameters}
Last but not least, we consider the effects of varying each of the following parameters --- $F$, $\kappa$ and $\mu$.  We find that varying $\kappa$ alters the final opinion distribution while varying the others do not.  This is consistent with the analysis above.  The reason is that by varying $\kappa$, we increase the fraction of similar agents and hence slow down the emergence of polarized opinion.  Therefore, the simulation result is very similar to the result shown in Fig.~\ref{fig:normal_vary_agreement}.  We do not show the corresponding histograms here to save space.

Surely, we may further consider the effects of $\bar{x}$ and $\sigma$.  However, we are not going to do so here.  Setting a non-zero $\bar{x}$ introduces initial bias on opinion distribution that may not be easy to justify.  Whereas if $\sigma \ll 1$, by rescaling the $X_f$, it is clear that the ratio $a / \sigma$ is a determining factor of the system dynamics.  As we have already investigated the effect of $a$ on the dynamics, this makes the study of $\sigma$ redundant.

\section{Reducing Opinion Polarization By Introducing Empathetic Agents}
\label{empathetic_agents}
Could we make the opinion distribution of the eventual states less polarized within the framework of our modified Axelrod model?  We would like to investigate possible interventions that could alleviate extreme opinion polarization.  As a general principle, we would like to make our interventions as small as possible, hopefully in the form of making small changes in our rules.  This is because changing human psychological behavior \emph{en masse} is not easy, at least in the short term.  Only these kinds of small changes have a real chance to occur in practice.  Here we propose to alter the behavior of a small fraction of agents.  Specifically, we fix a small portion $p_\text{s}$ of the agents randomly chosen among those whose number of neighbors in the B-A network rank top $100\rho_\text{s}$\% to be empathetic.  A empathetic agent always interacts with its neighbors using the rules of a similar agent.  One way to think of it is that each agent has a different agreement threshold, and that of a empathetic agent is arbitrarily large.  Their readiness to be influenced and persuaded by all kinds of opinions is the key to reaching a consensus in the population.

\begin{figure*}
    \centering
    \includegraphics[width = 0.9\textwidth]{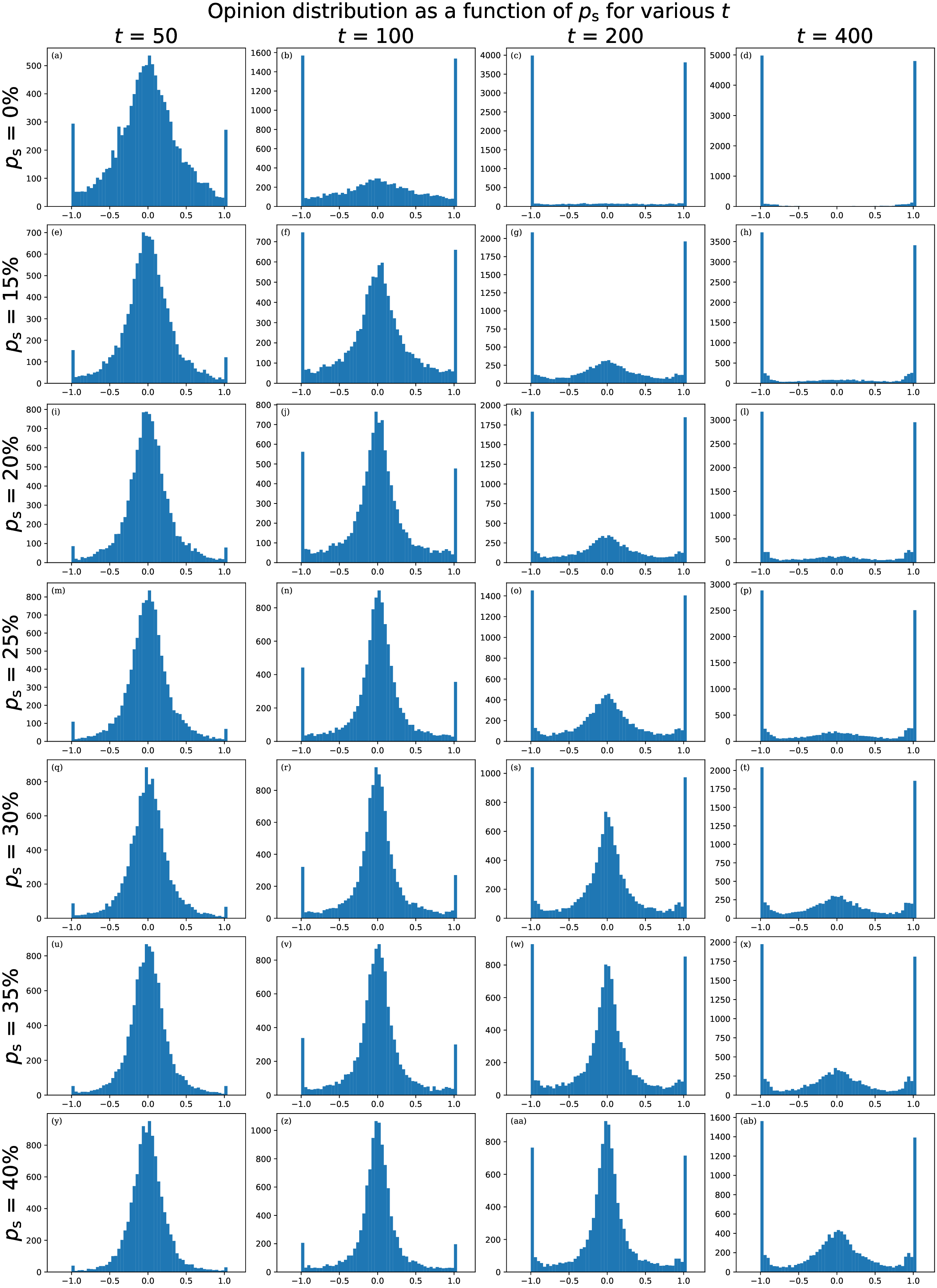}
    \caption{Evolution of opinion distribution for different portion of empathetic agents $p_\text{s}$ and mean number of interactions per agent $t$ when $\rho_\text{s} = 1$.  All other parameters are set to default values listed in Table~\ref{T:parameters}.}
    \label{fig:harm_vs_norm}
\end{figure*}

\begin{figure*}
    \centering
    \includegraphics[width = 0.9\textwidth]{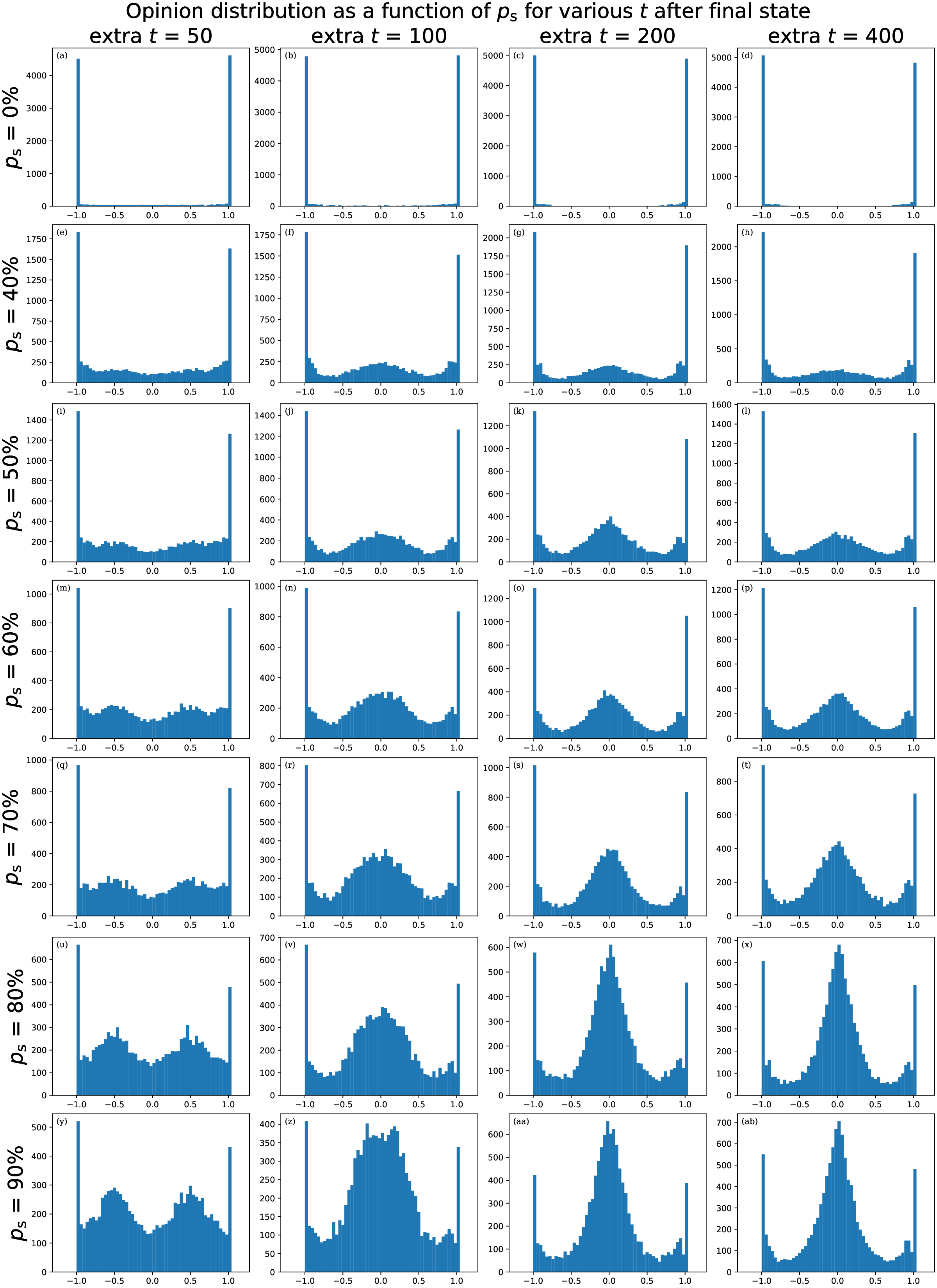}
    \caption{Evolution of opinion distribution for different values of $p_\text{s}$ after the system has almost reached a steady final state.  All parameters taken are the same as those in Fig.~\ref{fig:harm_vs_norm}.}
    \label{fig:harm_vs_norm_finalstate}
\end{figure*}

\begin{figure*}
    \centering
    \includegraphics[width = 0.9\textwidth]{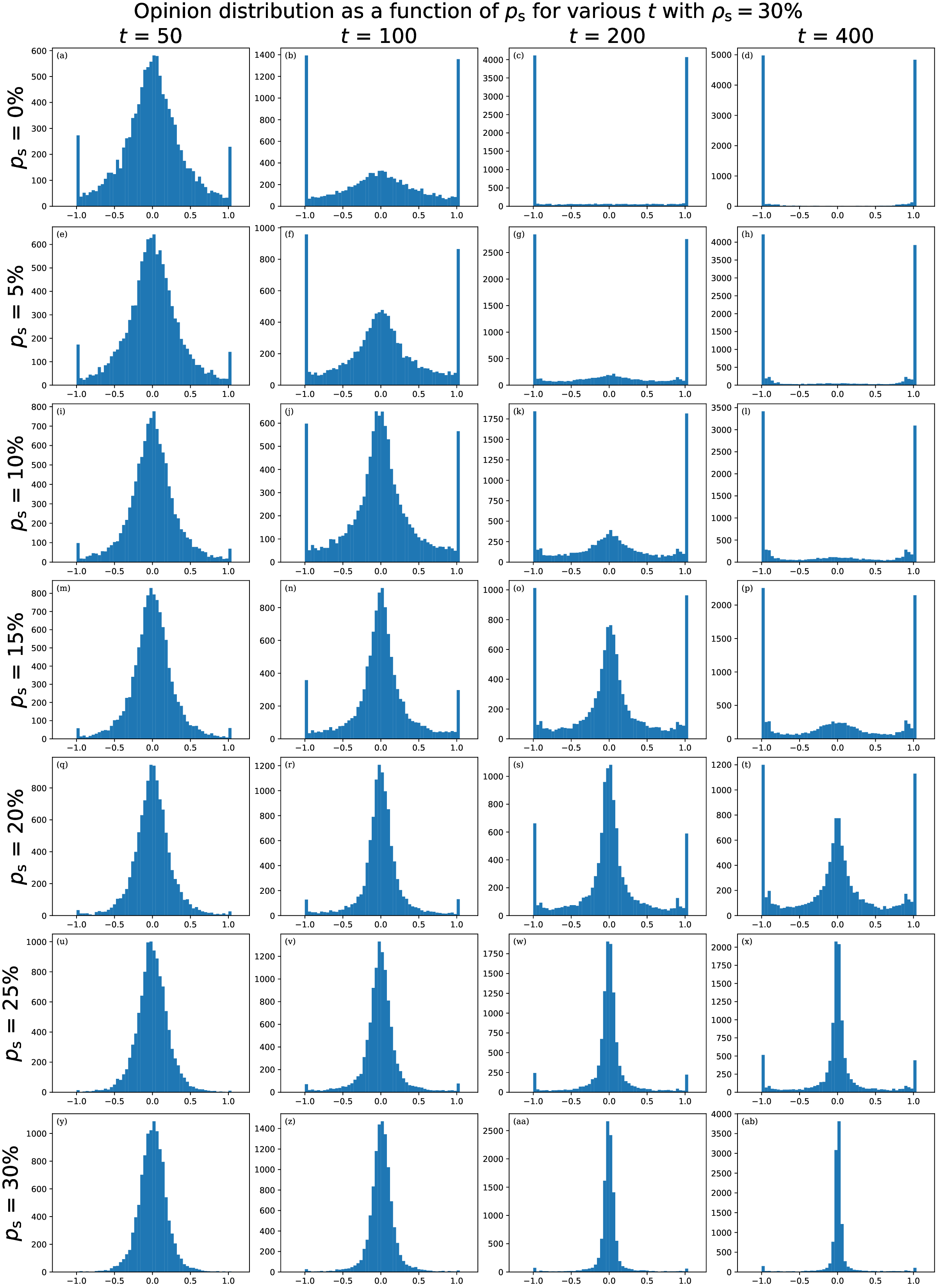}
    \caption{Evolution of opinion distribution for different portion of empathetic agents $p_\text{s}$ and mean number of interactions per agent $t$ when $\rho_\text{s} = 0.3$.  All other parameters are set to default values listed in Table~\ref{T:parameters}.}
    \label{fig:harm_vs_norm_high_connections}
\end{figure*}

\begin{figure*}
    \centering
    \includegraphics[width = 0.9\textwidth]{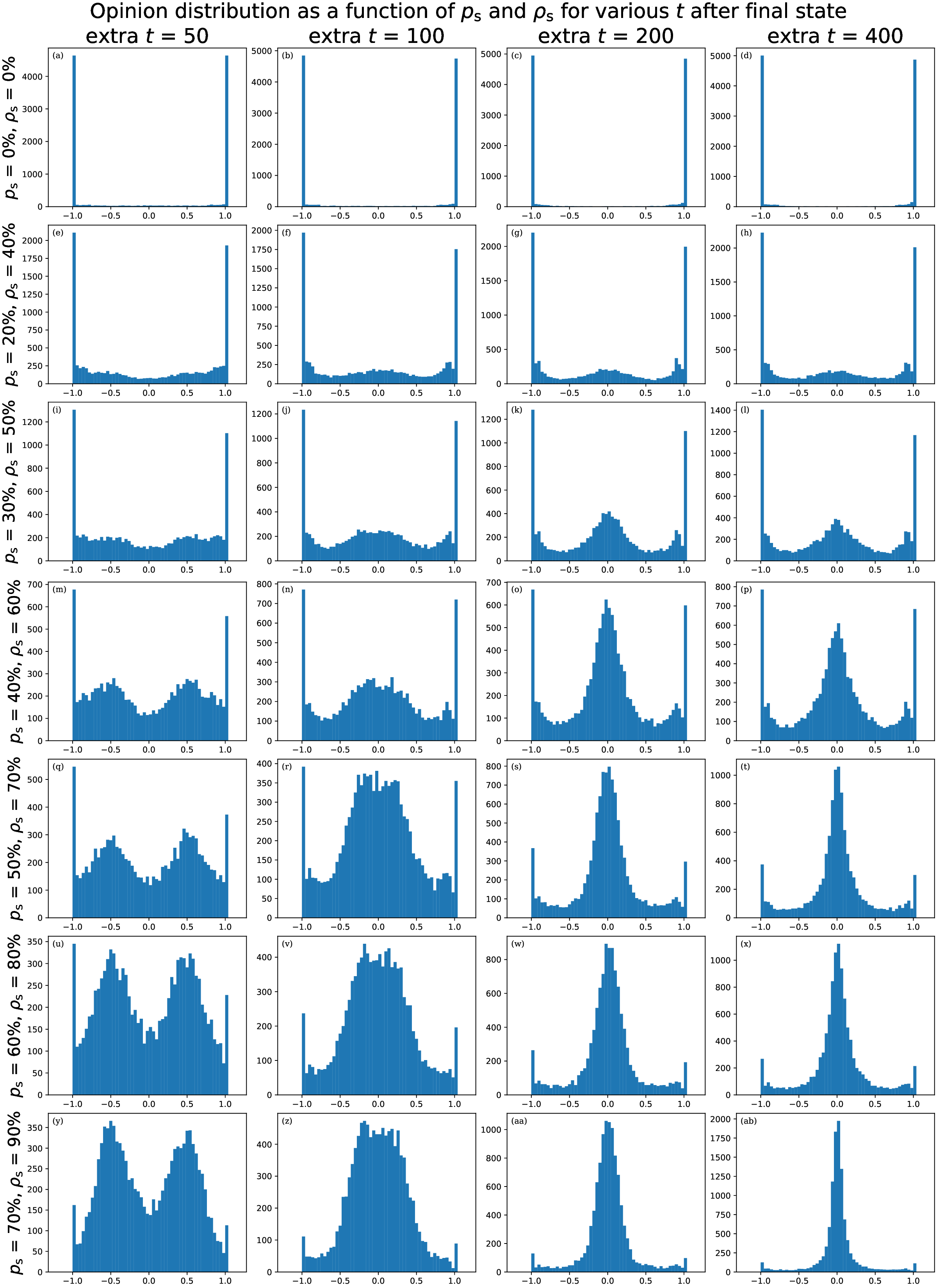}
    \caption{Evolution of opinion distribution for different values of $p_\text{s}$ after the system has almost reached a steady final state.  All parameters taken are the same as those in Fig.~\ref{fig:harm_vs_norm_high_connections}.}
    \label{fig:harm_vs_norm_finalstate_high_connections}
\end{figure*}

\subsection{The Case Of Unbiased Empathetic Agent Selection}
We first consider the case when $\rho_\text{s} = 1$ so that each agent has equally likely chance of chosen to be empathetic.
\label{Subsec:empathetic_1}
Here, we investigate the opinion evolution of systems with these empathetic agents in two ways.  Fig.~\ref{fig:harm_vs_norm} depicts the opinion distribution evolution when empathetic agents are assigned in the beginning.  As expected, the higher the value of $p_\text{s}$, the smaller the deviation in the attitudes of all agents.  Although a sizable number of extremists is still present, a significant fraction of moderates remain.  Nevertheless, this effect is significant only when $p_\text{s} \gtrsim 35\%$.  Such a high percentage of empathetic agents is very hard if not impossible to reach in reality.

A more pressing issue is how to reduce society polarization in an already polarized society.  Therefore, we study what happens if we change a fixed portion of agents from ``normal'' to empathetic \emph{after} the society has evolved almost to its final highly polarized state.  More precisely, we first let the system evolve with $t = 200$ using parameters listed in Table~\ref{T:parameters} so that it has almost equilibrated.  Then, we will randomly pick and fix $p_\text{s}$ of the agents and make them empathetic. Finally, we further evolve the system for an additional $t$ up to $400$.  From Fig.~\ref{fig:harm_vs_norm_finalstate}, we observe that one can significantly reduce opinion polarization only when $p_\text{s} \gtrsim 40\%$, a goal that is even harder to attain.

\subsection{The Case When Empathetic Agents Are Picked From The High Degree Nodes}
In view of the ineffectiveness of picking empathetic agents randomly from the population, we consider the case of choosing them only from highly connected agents.  In the real world, this corresponds to the case that a sizable number of social media influencers are empathetic.  First we redo our simulations for the case of assigning empathetic agents right from the start by fixing $\rho_\text{s} = 0.3$ as an illustration.  As shown in Fig.~\ref{fig:harm_vs_norm_high_connections}, a significant fraction of moderates remains when $p_\text{s} \gtrsim 20\%$ (which is much less than the critical value of $p_s \approx 35\%$ for the case when $\rho_\text{s} = 1$).  The much more interesting case is when selected ``normal'' agents turn empathetic after the system has evolved with $t = 200$.  Fig.~\ref{fig:harm_vs_norm_finalstate_high_connections} depicts that system becomes much less polarized if $\rho_\text{s} = 50\%$ and $p_\text{s} \gtrsim 30\%$.  This corresponds to changing the behavior of about $60\%$ of those agents whose connectivity is above average.  Furthermore, if picking $\rho_\text{s} = 40\%$ and $p_\text{s} = 60\%$, moderate opinion agents domain the system in the end.  Compared to the case of randomly assigning empathetic agents, these strategies are much more viable although it is still highly challenging to implement.

\section{Conclusions}
In this paper, we have identified deficiencies in the Axelrod model of opinion dynamics and proposed a modified one as an attempt to build a more sophisticated model to describe opinion dynamics in human society.  Major modifications include using scale-free networks replacing the static two-dimensional lattice grid together with a new set of interaction rules for similar and dissimilar agents.
Our simulation results show that over essentially all reasonable parameters agents' opinions on all features eventually become more and more polarized.  Besides, one or two dominating opinion clusters are formed together with numerous number of small opinion clusters.  This is not unlike the big opinion differences between those living in cities and in rural areas as well as the emergence of political extremism due to distrust of the elites in the real world.
We also discovered scaling in the distribution of the relative opinion cluster size $n(s,N)/N$.  For B-A network, it follows a power law when $s \lesssim 10$ and switches to exponential decay when $15 \lesssim s \lesssim 35$.
Finally, we tried to reduce this extreme opinion polarization by introducing empathetic agents that are open to talking and actively interacting with their neighbors.
Their presence reduces opinion polarization but only when they appear \emph{en masse}.  Our study suggests that one effective and relatively economical but highly challenging way to lessen the difference in an already highly polarized opinion population in the real world is to change the behavior of a significant portion of social media influencers making them more empathetic.

Our model is flexible and can be modified to consider more realistic cases, even including evolution on the network.  Our research is a small first step toward finding viable methods to understand opinion dynamics and reduce opinion polarization in our contemporary society.

\acknowledgments
XZ was partially supported by The Innovation and Technology Scholarship of the
Hong Kong Federation of Youth Groups.

\bibliographystyle{apsrev4-2}
\bibliography{socphys3.3.bib}

\begin{thebibliography}{52}%
\makeatletter
\providecommand \@ifxundefined [1]{%
 \@ifx{#1\undefined}
}%
\providecommand \@ifnum [1]{%
 \ifnum #1\expandafter \@firstoftwo
 \else \expandafter \@secondoftwo
 \fi
}%
\providecommand \@ifx [1]{%
 \ifx #1\expandafter \@firstoftwo
 \else \expandafter \@secondoftwo
 \fi
}%
\providecommand \natexlab [1]{#1}%
\providecommand \enquote  [1]{``#1''}%
\providecommand \bibnamefont  [1]{#1}%
\providecommand \bibfnamefont [1]{#1}%
\providecommand \citenamefont [1]{#1}%
\providecommand \href@noop [0]{\@secondoftwo}%
\providecommand \href [0]{\begingroup \@sanitize@url \@href}%
\providecommand \@href[1]{\@@startlink{#1}\@@href}%
\providecommand \@@href[1]{\endgroup#1\@@endlink}%
\providecommand \@sanitize@url [0]{\catcode `\\12\catcode `\$12\catcode
  `\&12\catcode `\#12\catcode `\^12\catcode `\_12\catcode `\%12\relax}%
\providecommand \@@startlink[1]{}%
\providecommand \@@endlink[0]{}%
\providecommand \url  [0]{\begingroup\@sanitize@url \@url }%
\providecommand \@url [1]{\endgroup\@href {#1}{\urlprefix }}%
\providecommand \urlprefix  [0]{URL }%
\providecommand \Eprint [0]{\href }%
\providecommand \doibase [0]{https://doi.org/}%
\providecommand \selectlanguage [0]{\@gobble}%
\providecommand \bibinfo  [0]{\@secondoftwo}%
\providecommand \bibfield  [0]{\@secondoftwo}%
\providecommand \translation [1]{[#1]}%
\providecommand \BibitemOpen [0]{}%
\providecommand \bibitemStop [0]{}%
\providecommand \bibitemNoStop [0]{.\EOS\space}%
\providecommand \EOS [0]{\spacefactor3000\relax}%
\providecommand \BibitemShut  [1]{\csname bibitem#1\endcsname}%
\let\auto@bib@innerbib\@empty
\bibitem [{\citenamefont {Jusup}\ \emph {et~al.}(2022)\citenamefont {Jusup},
  \citenamefont {Holme}, \citenamefont {Kanazawa}, \citenamefont {Takayasu},
  \citenamefont {Romić}, \citenamefont {Wang}, \citenamefont {Geček},
  \citenamefont {Lipić}, \citenamefont {Podobnik}, \citenamefont {Wang},
  \citenamefont {Luo}, \citenamefont {Klanjšček}, \citenamefont {Fan},
  \citenamefont {Boccaletti},\ and\ \citenamefont
  {Perc}}]{Sociophysics_review}%
  \BibitemOpen
  \bibfield  {author} {\bibinfo {author} {\bibfnamefont {M.}~\bibnamefont
  {Jusup}}, \bibinfo {author} {\bibfnamefont {P.}~\bibnamefont {Holme}},
  \bibinfo {author} {\bibfnamefont {K.}~\bibnamefont {Kanazawa}}, \bibinfo
  {author} {\bibfnamefont {M.}~\bibnamefont {Takayasu}}, \bibinfo {author}
  {\bibfnamefont {I.}~\bibnamefont {Romić}}, \bibinfo {author} {\bibfnamefont
  {Z.}~\bibnamefont {Wang}}, \bibinfo {author} {\bibfnamefont {S.}~\bibnamefont
  {Geček}}, \bibinfo {author} {\bibfnamefont {T.}~\bibnamefont {Lipić}},
  \bibinfo {author} {\bibfnamefont {B.}~\bibnamefont {Podobnik}}, \bibinfo
  {author} {\bibfnamefont {L.}~\bibnamefont {Wang}}, \bibinfo {author}
  {\bibfnamefont {W.}~\bibnamefont {Luo}}, \bibinfo {author} {\bibfnamefont
  {T.}~\bibnamefont {Klanjšček}}, \bibinfo {author} {\bibfnamefont
  {J.}~\bibnamefont {Fan}}, \bibinfo {author} {\bibfnamefont {S.}~\bibnamefont
  {Boccaletti}},\ and\ \bibinfo {author} {\bibfnamefont {M.}~\bibnamefont
  {Perc}},\ }\href
  {https://www.sciencedirect.com/science/article/pii/S037015732100404X}
  {\bibfield  {journal} {\bibinfo  {journal} {Phys. Rep.}\ }\textbf {\bibinfo
  {volume} {948}},\ \bibinfo {pages} {1} (\bibinfo {year} {2022})}\BibitemShut
  {NoStop}%
\bibitem [{\citenamefont {Sznajd-Weron}\ and\ \citenamefont
  {Sznajd}(2000)}]{Sznajd_model1}%
  \BibitemOpen
  \bibfield  {author} {\bibinfo {author} {\bibfnamefont {K.}~\bibnamefont
  {Sznajd-Weron}}\ and\ \bibinfo {author} {\bibfnamefont {J.}~\bibnamefont
  {Sznajd}},\ }\href@noop {} {\bibfield  {journal} {\bibinfo  {journal} {Int.
  J. Mod. Phys. C}\ }\textbf {\bibinfo {volume} {11}},\ \bibinfo {pages} {1157}
  (\bibinfo {year} {2000})}\BibitemShut {NoStop}%
\bibitem [{\citenamefont {Sznajd-Weron}(2005)}]{Sznajd_model2}%
  \BibitemOpen
  \bibfield  {author} {\bibinfo {author} {\bibfnamefont {K.}~\bibnamefont
  {Sznajd-Weron}},\ }\href@noop {} {\bibfield  {journal} {\bibinfo  {journal}
  {Acta Phys. Pol. B}\ }\textbf {\bibinfo {volume} {36}},\ \bibinfo {pages}
  {2537} (\bibinfo {year} {2005})}\BibitemShut {NoStop}%
\bibitem [{\citenamefont {Hegselmann}\ and\ \citenamefont
  {Krause}(2002)}]{Hegselmann_Krause_model}%
  \BibitemOpen
  \bibfield  {author} {\bibinfo {author} {\bibfnamefont {R.}~\bibnamefont
  {Hegselmann}}\ and\ \bibinfo {author} {\bibfnamefont {U.}~\bibnamefont
  {Krause}},\ }\href {https://www.jasss.org/5/3/2.html} {\bibfield  {journal}
  {\bibinfo  {journal} {J. Artifical Societies and Social Simulation}\ }\textbf
  {\bibinfo {volume} {5}},\ \bibinfo {pages} {2} (\bibinfo {year} {2002})},\
  \Eprint {https://arxiv.org/abs/https://www.jasss.org/5/3/2.html}
  {https://www.jasss.org/5/3/2.html} \BibitemShut {NoStop}%
\bibitem [{\citenamefont {Galam}(2008)}]{Galam_models}%
  \BibitemOpen
  \bibfield  {author} {\bibinfo {author} {\bibfnamefont {S.}~\bibnamefont
  {Galam}},\ }\href@noop {} {\bibfield  {journal} {\bibinfo  {journal} {Int. J.
  Mod. Phys. C}\ }\textbf {\bibinfo {volume} {19}},\ \bibinfo {pages} {409}
  (\bibinfo {year} {2008})}\BibitemShut {NoStop}%
\bibitem [{\citenamefont {Axelrod}(1997)}]{Axelrod1997}%
  \BibitemOpen
  \bibfield  {author} {\bibinfo {author} {\bibfnamefont {R.}~\bibnamefont
  {Axelrod}},\ }\href {http://www.jstor.org/stable/174371} {\bibfield
  {journal} {\bibinfo  {journal} {J. Conflict Resolut.}\ }\textbf {\bibinfo
  {volume} {41}},\ \bibinfo {pages} {203} (\bibinfo {year} {1997})}\BibitemShut
  {NoStop}%
\bibitem [{\citenamefont {Deffuant}\ \emph {et~al.}(2000)\citenamefont
  {Deffuant}, \citenamefont {Neau}, \citenamefont {Amblard},\ and\
  \citenamefont {Weisbuch}}]{Deffuant2000}%
  \BibitemOpen
  \bibfield  {author} {\bibinfo {author} {\bibfnamefont {G.}~\bibnamefont
  {Deffuant}}, \bibinfo {author} {\bibfnamefont {D.}~\bibnamefont {Neau}},
  \bibinfo {author} {\bibfnamefont {F.}~\bibnamefont {Amblard}},\ and\ \bibinfo
  {author} {\bibfnamefont {G.}~\bibnamefont {Weisbuch}},\ }\href@noop {}
  {\bibfield  {journal} {\bibinfo  {journal} {Adv. Complex Sys.}\ }\textbf
  {\bibinfo {volume} {3}},\ \bibinfo {pages} {87} (\bibinfo {year}
  {2000})}\BibitemShut {NoStop}%
\bibitem [{\citenamefont {Weisbuch}\ \emph {et~al.}(2002)\citenamefont
  {Weisbuch}, \citenamefont {Deffuant}, \citenamefont {Amblard},\ and\
  \citenamefont {Nadal}}]{Weisbuch2002}%
  \BibitemOpen
  \bibfield  {author} {\bibinfo {author} {\bibfnamefont {G.}~\bibnamefont
  {Weisbuch}}, \bibinfo {author} {\bibfnamefont {G.}~\bibnamefont {Deffuant}},
  \bibinfo {author} {\bibfnamefont {F.}~\bibnamefont {Amblard}},\ and\ \bibinfo
  {author} {\bibfnamefont {J.-P.}\ \bibnamefont {Nadal}},\ }\href@noop {}
  {\bibfield  {journal} {\bibinfo  {journal} {Complexity}\ }\textbf {\bibinfo
  {volume} {7}},\ \bibinfo {pages} {55} (\bibinfo {year} {2002})}\BibitemShut
  {NoStop}%
\bibitem [{\citenamefont {S{\^i}rbu}\ \emph {et~al.}(2016)\citenamefont
  {S{\^i}rbu}, \citenamefont {Loreto}, \citenamefont {Servedio},\ and\
  \citenamefont {Tria}}]{Sociophysicsmodel1}%
  \BibitemOpen
  \bibfield  {author} {\bibinfo {author} {\bibfnamefont {A.}~\bibnamefont
  {S{\^i}rbu}}, \bibinfo {author} {\bibfnamefont {V.}~\bibnamefont {Loreto}},
  \bibinfo {author} {\bibfnamefont {V.~D.~P.}\ \bibnamefont {Servedio}},\ and\
  \bibinfo {author} {\bibfnamefont {F.}~\bibnamefont {Tria}},\ }in\ \href
  {{https://doi.org/10.1007/978-3-319-25658-0\_17}} {\emph {\bibinfo
  {booktitle} {Participatory Sensing, Opinions and Collective Awareness}}},\
  \bibinfo {editor} {edited by\ \bibinfo {editor} {\bibfnamefont
  {V.}~\bibnamefont {Loreto}}, \bibinfo {editor} {\bibfnamefont
  {M.}~\bibnamefont {Haklay}}, \bibinfo {editor} {\bibfnamefont
  {A.}~\bibnamefont {Hotho}}, \bibinfo {editor} {\bibfnamefont {V.~D.}\
  \bibnamefont {Servedio}}, \bibinfo {editor} {\bibfnamefont {G.}~\bibnamefont
  {Stumme}}, \bibinfo {editor} {\bibfnamefont {J.}~\bibnamefont {Theunis}},\
  and\ \bibinfo {editor} {\bibfnamefont {F.}~\bibnamefont {Tria}}}\ (\bibinfo
  {publisher} {Springer International Publishing},\ \bibinfo {address} {Cham},\
  \bibinfo {year} {2016})\ pp.\ \bibinfo {pages} {363--401}\BibitemShut
  {NoStop}%
\bibitem [{\citenamefont {Peralta}\ \emph {et~al.}(2025)\citenamefont
  {Peralta}, \citenamefont {Kertész},\ and\ \citenamefont
  {Iñiguez}}]{Sociophysicsmodel2}%
  \BibitemOpen
  \bibfield  {author} {\bibinfo {author} {\bibfnamefont {A.~F.}\ \bibnamefont
  {Peralta}}, \bibinfo {author} {\bibfnamefont {J.}~\bibnamefont {Kertész}},\
  and\ \bibinfo {author} {\bibfnamefont {G.}~\bibnamefont {Iñiguez}},\
  }\bibinfo {title} {Opinion dynamics in social networks: From models to
  data},\ in\ \href@noop {} {\emph {\bibinfo {booktitle} {Handbook Of
  Computational Social Science}}},\ \bibinfo {editor} {edited by\ \bibinfo
  {editor} {\bibfnamefont {T.}~\bibnamefont {Yasseri}}}\ (\bibinfo  {publisher}
  {Elgaronline},\ \bibinfo {year} {2025})\ pp.\ \bibinfo {pages}
  {384--406}\BibitemShut {NoStop}%
\bibitem [{\citenamefont {Liu}\ \emph {et~al.}(2022)\citenamefont {Liu},
  \citenamefont {He}, \citenamefont {Qiu},\ and\ \citenamefont
  {He}}]{Sociophysicsmodel3}%
  \BibitemOpen
  \bibfield  {author} {\bibinfo {author} {\bibfnamefont {J.}~\bibnamefont
  {Liu}}, \bibinfo {author} {\bibfnamefont {J.}~\bibnamefont {He}}, \bibinfo
  {author} {\bibfnamefont {Z.}~\bibnamefont {Qiu}},\ and\ \bibinfo {author}
  {\bibfnamefont {S.}~\bibnamefont {He}},\ }\bibfield  {journal} {\bibinfo
  {journal} {Front. Phys.}\ }\textbf {\bibinfo {volume} {10}},\ \href
  {https://doi.org/10.3389/fphy.2022.1042900} {10.3389/fphy.2022.1042900}
  (\bibinfo {year} {2022})\BibitemShut {NoStop}%
\bibitem [{\citenamefont {Helfmann}\ \emph {et~al.}(2023)\citenamefont
  {Helfmann}, \citenamefont {Djurdjevac~Conrad}, \citenamefont
  {Lorenz-Spreen},\ and\ \citenamefont {Sch{\"u}tte}}]{Sociophysicsmodel4}%
  \BibitemOpen
  \bibfield  {author} {\bibinfo {author} {\bibfnamefont {L.}~\bibnamefont
  {Helfmann}}, \bibinfo {author} {\bibfnamefont {N.}~\bibnamefont
  {Djurdjevac~Conrad}}, \bibinfo {author} {\bibfnamefont {P.}~\bibnamefont
  {Lorenz-Spreen}},\ and\ \bibinfo {author} {\bibfnamefont {C.}~\bibnamefont
  {Sch{\"u}tte}},\ }\href@noop {} {\bibfield  {journal} {\bibinfo  {journal}
  {Sci. Rep.}\ }\textbf {\bibinfo {volume} {13}},\ \bibinfo {pages} {19375}
  (\bibinfo {year} {2023})}\BibitemShut {NoStop}%
\bibitem [{\citenamefont {Chau}\ \emph {et~al.}(2014)\citenamefont {Chau},
  \citenamefont {Wong}, \citenamefont {Chow},\ and\ \citenamefont
  {Fung}}]{Chau2014}%
  \BibitemOpen
  \bibfield  {author} {\bibinfo {author} {\bibfnamefont {H.}~\bibnamefont
  {Chau}}, \bibinfo {author} {\bibfnamefont {C.}~\bibnamefont {Wong}}, \bibinfo
  {author} {\bibfnamefont {F.}~\bibnamefont {Chow}},\ and\ \bibinfo {author}
  {\bibfnamefont {C.-H.~F.}\ \bibnamefont {Fung}},\ }\href
  {https://www.sciencedirect.com/science/article/pii/S0378437114006682}
  {\bibfield  {journal} {\bibinfo  {journal} {Physica A}\ }\textbf {\bibinfo
  {volume} {415}},\ \bibinfo {pages} {133} (\bibinfo {year}
  {2014})}\BibitemShut {NoStop}%
\bibitem [{\citenamefont {Castellano}\ \emph {et~al.}(2000)\citenamefont
  {Castellano}, \citenamefont {Marsili},\ and\ \citenamefont
  {Vespignani}}]{CMV00}%
  \BibitemOpen
  \bibfield  {author} {\bibinfo {author} {\bibfnamefont {C.}~\bibnamefont
  {Castellano}}, \bibinfo {author} {\bibfnamefont {M.}~\bibnamefont
  {Marsili}},\ and\ \bibinfo {author} {\bibfnamefont {A.}~\bibnamefont
  {Vespignani}},\ }\href@noop {} {\bibfield  {journal} {\bibinfo  {journal}
  {Phys. Rev. Lett.}\ }\textbf {\bibinfo {volume} {85}},\ \bibinfo {pages}
  {3536} (\bibinfo {year} {2000})}\BibitemShut {NoStop}%
\bibitem [{\citenamefont {Klemm}\ \emph {et~al.}(2003)\citenamefont {Klemm},
  \citenamefont {Eguiluz}, \citenamefont {Toral},\ and\ \citenamefont
  {Miguel}}]{KETM03}%
  \BibitemOpen
  \bibfield  {author} {\bibinfo {author} {\bibfnamefont {K.}~\bibnamefont
  {Klemm}}, \bibinfo {author} {\bibfnamefont {V.~M.}\ \bibnamefont {Eguiluz}},
  \bibinfo {author} {\bibfnamefont {R.}~\bibnamefont {Toral}},\ and\ \bibinfo
  {author} {\bibfnamefont {M.~S.}\ \bibnamefont {Miguel}},\ }\href@noop {}
  {\bibfield  {journal} {\bibinfo  {journal} {Phys. Rev. E}\ }\textbf {\bibinfo
  {volume} {67}},\ \bibinfo {pages} {045101(R)} (\bibinfo {year}
  {2003})}\BibitemShut {NoStop}%
\bibitem [{\citenamefont {Ashraf}\ and\ \citenamefont
  {Galor}(2013)}]{cultural_fragmentation1}%
  \BibitemOpen
  \bibfield  {author} {\bibinfo {author} {\bibfnamefont {Q.}~\bibnamefont
  {Ashraf}}\ and\ \bibinfo {author} {\bibfnamefont {O.}~\bibnamefont {Galor}},\
  }\href {https://doi.org/10.1257/aer.103.3.528} {\bibfield  {journal}
  {\bibinfo  {journal} {Am. Econ. Rev.}\ }\textbf {\bibinfo {volume} {103}},\
  \bibinfo {pages} {528} (\bibinfo {year} {2013})}\BibitemShut {NoStop}%
\bibitem [{\citenamefont {Dandekar}\ \emph {et~al.}(2013)\citenamefont
  {Dandekar}, \citenamefont {Goel},\ and\ \citenamefont
  {Lee}}]{political_diverse1}%
  \BibitemOpen
  \bibfield  {author} {\bibinfo {author} {\bibfnamefont {P.}~\bibnamefont
  {Dandekar}}, \bibinfo {author} {\bibfnamefont {A.}~\bibnamefont {Goel}},\
  and\ \bibinfo {author} {\bibfnamefont {D.~T.}\ \bibnamefont {Lee}},\
  }\href@noop {} {\bibfield  {journal} {\bibinfo  {journal} {Proc. Natl. Acad.
  Sci. U.S.A.}\ }\textbf {\bibinfo {volume} {110}},\ \bibinfo {pages} {5791}
  (\bibinfo {year} {2013})}\BibitemShut {NoStop}%
\bibitem [{\citenamefont {Koudenburg}\ and\ \citenamefont
  {Kashima}(2022)}]{political_diverse2}%
  \BibitemOpen
  \bibfield  {author} {\bibinfo {author} {\bibfnamefont {N.}~\bibnamefont
  {Koudenburg}}\ and\ \bibinfo {author} {\bibfnamefont {Y.}~\bibnamefont
  {Kashima}},\ }\href@noop {} {\bibfield  {journal} {\bibinfo  {journal} {Pers.
  Soc. Psychol. Bull.}\ }\textbf {\bibinfo {volume} {48}},\ \bibinfo {pages}
  {1068} (\bibinfo {year} {2022})}\BibitemShut {NoStop}%
\bibitem [{\citenamefont {Ramos}\ \emph {et~al.}(2015)\citenamefont {Ramos},
  \citenamefont {Shao}, \citenamefont {Reis}, \citenamefont {Anteneodo},
  \citenamefont {Andrade}, \citenamefont {Havlin},\ and\ \citenamefont
  {Makse}}]{political_diverse3}%
  \BibitemOpen
  \bibfield  {author} {\bibinfo {author} {\bibfnamefont {M.}~\bibnamefont
  {Ramos}}, \bibinfo {author} {\bibfnamefont {J.}~\bibnamefont {Shao}},
  \bibinfo {author} {\bibfnamefont {S.~D.~S.}\ \bibnamefont {Reis}}, \bibinfo
  {author} {\bibfnamefont {C.}~\bibnamefont {Anteneodo}}, \bibinfo {author}
  {\bibfnamefont {J.~S.}\ \bibnamefont {Andrade}}, \bibinfo {author}
  {\bibfnamefont {S.}~\bibnamefont {Havlin}},\ and\ \bibinfo {author}
  {\bibfnamefont {H.~A.}\ \bibnamefont {Makse}},\ }\href@noop {} {\bibfield
  {journal} {\bibinfo  {journal} {Sci. Rep.}\ }\textbf {\bibinfo {volume}
  {5}},\ \bibinfo {pages} {10032} (\bibinfo {year} {2015})}\BibitemShut
  {NoStop}%
\bibitem [{\citenamefont {Devauchelle}\ \emph {et~al.}(2024)\citenamefont
  {Devauchelle}, \citenamefont {Szymczak},\ and\ \citenamefont
  {Nowakowski}}]{Devauchelle2024}%
  \BibitemOpen
  \bibfield  {author} {\bibinfo {author} {\bibfnamefont {O.}~\bibnamefont
  {Devauchelle}}, \bibinfo {author} {\bibfnamefont {P.}~\bibnamefont
  {Szymczak}},\ and\ \bibinfo {author} {\bibfnamefont {P.}~\bibnamefont
  {Nowakowski}},\ }\href {https://link.aps.org/doi/10.1103/PhysRevE.109.044106}
  {\bibfield  {journal} {\bibinfo  {journal} {Phys. Rev. E}\ }\textbf {\bibinfo
  {volume} {109}},\ \bibinfo {pages} {044106} (\bibinfo {year}
  {2024})}\BibitemShut {NoStop}%
\bibitem [{\citenamefont {MacCarron}\ \emph {et~al.}(2020)\citenamefont
  {MacCarron}, \citenamefont {Maher}, \citenamefont {Fennell}, \citenamefont
  {Burke}, \citenamefont {Gleeson}, \citenamefont {Durrheim},\ and\
  \citenamefont {Quayle}}]{spreading_culture}%
  \BibitemOpen
  \bibfield  {author} {\bibinfo {author} {\bibfnamefont {P.}~\bibnamefont
  {MacCarron}}, \bibinfo {author} {\bibfnamefont {P.~J.}\ \bibnamefont
  {Maher}}, \bibinfo {author} {\bibfnamefont {S.}~\bibnamefont {Fennell}},
  \bibinfo {author} {\bibfnamefont {K.}~\bibnamefont {Burke}}, \bibinfo
  {author} {\bibfnamefont {J.~P.}\ \bibnamefont {Gleeson}}, \bibinfo {author}
  {\bibfnamefont {K.}~\bibnamefont {Durrheim}},\ and\ \bibinfo {author}
  {\bibfnamefont {M.}~\bibnamefont {Quayle}},\ }\href
  {https://doi.org/10.1371/journal.pone.0233995} {\bibfield  {journal}
  {\bibinfo  {journal} {PLOS ONE}\ }\textbf {\bibinfo {volume} {15}},\ \bibinfo
  {pages} {e0233995} (\bibinfo {year} {2020})}\BibitemShut {NoStop}%
\bibitem [{\citenamefont {Radilo-D{\'i}az}\ \emph {et~al.}(2009)\citenamefont
  {Radilo-D{\'i}az}, \citenamefont {P{\'e}rez},\ and\ \citenamefont {del
  Castillo-Mussot}}]{RD09}%
  \BibitemOpen
  \bibfield  {author} {\bibinfo {author} {\bibfnamefont {A.}~\bibnamefont
  {Radilo-D{\'i}az}}, \bibinfo {author} {\bibfnamefont {L.~A.}\ \bibnamefont
  {P{\'e}rez}},\ and\ \bibinfo {author} {\bibfnamefont {M.}~\bibnamefont {del
  Castillo-Mussot}},\ }\href@noop {} {\bibfield  {journal} {\bibinfo  {journal}
  {Phys. Rev. E}\ }\textbf {\bibinfo {volume} {80}},\ \bibinfo {pages} {066107}
  (\bibinfo {year} {2009})}\BibitemShut {NoStop}%
\bibitem [{\citenamefont {Battiston}\ \emph {et~al.}(2017)\citenamefont
  {Battiston}, \citenamefont {Nicosia}, \citenamefont {Latora},\ and\
  \citenamefont {Miguel}}]{Battiston2017}%
  \BibitemOpen
  \bibfield  {author} {\bibinfo {author} {\bibfnamefont {F.}~\bibnamefont
  {Battiston}}, \bibinfo {author} {\bibfnamefont {V.}~\bibnamefont {Nicosia}},
  \bibinfo {author} {\bibfnamefont {V.}~\bibnamefont {Latora}},\ and\ \bibinfo
  {author} {\bibfnamefont {M.~S.}\ \bibnamefont {Miguel}},\ }\href
  {https://doi.org/10.1038/s41598-017-02040-4} {\bibfield  {journal} {\bibinfo
  {journal} {Sci. Rep.}\ }\textbf {\bibinfo {volume} {7}},\ \bibinfo {pages}
  {1809} (\bibinfo {year} {2017})}\BibitemShut {NoStop}%
\bibitem [{\citenamefont {T{\"o}rnberg}(2022)}]{T22}%
  \BibitemOpen
  \bibfield  {author} {\bibinfo {author} {\bibfnamefont {P.}~\bibnamefont
  {T{\"o}rnberg}},\ }\href@noop {} {\bibfield  {journal} {\bibinfo  {journal}
  {Proc. Natl. Acad. Sci. U.S.A.}\ }\textbf {\bibinfo {volume} {119}},\
  \bibinfo {pages} {e2207159119} (\bibinfo {year} {2022})}\BibitemShut
  {NoStop}%
\bibitem [{\citenamefont {Amorim}(2014)}]{Amorim2014}%
  \BibitemOpen
  \bibfield  {author} {\bibinfo {author} {\bibfnamefont {C.~S.}\ \bibnamefont
  {Amorim}},\ }\href@noop {} {\bibinfo {title} {Bourdieu dynamics of fields
  from a modified {A}xelrod model}} (\bibinfo {year} {2014}),\ \Eprint
  {https://arxiv.org/abs/1407.3479} {arXiv:1407.3479 [physics.soc-ph]}
  \BibitemShut {NoStop}%
\bibitem [{\citenamefont {Albert}\ and\ \citenamefont
  {Barab\'asi}(2002)}]{BAnetwork}%
  \BibitemOpen
  \bibfield  {author} {\bibinfo {author} {\bibfnamefont {R.}~\bibnamefont
  {Albert}}\ and\ \bibinfo {author} {\bibfnamefont {A.-L.}\ \bibnamefont
  {Barab\'asi}},\ }\href@noop {} {\bibfield  {journal} {\bibinfo  {journal}
  {Rev. Mod. Phys.}\ }\textbf {\bibinfo {volume} {74}},\ \bibinfo {pages} {47}
  (\bibinfo {year} {2002})}\BibitemShut {NoStop}%
\bibitem [{\citenamefont {Jager}\ and\ \citenamefont
  {Amblard}(2005)}]{JAmodel}%
  \BibitemOpen
  \bibfield  {author} {\bibinfo {author} {\bibfnamefont {W.}~\bibnamefont
  {Jager}}\ and\ \bibinfo {author} {\bibfnamefont {F.}~\bibnamefont
  {Amblard}},\ }\href {https://doi.org/10.1007/s10588-005-6282-2} {\bibfield
  {journal} {\bibinfo  {journal} {Comput. Math. Organ. Theo.}\ }\textbf
  {\bibinfo {volume} {10}},\ \bibinfo {pages} {295} (\bibinfo {year}
  {2005})}\BibitemShut {NoStop}%
\bibitem [{\citenamefont {Bartolozzi}\ \emph {et~al.}(2005)\citenamefont
  {Bartolozzi}, \citenamefont {Leinweber},\ and\ \citenamefont
  {Thomas}}]{BLT05}%
  \BibitemOpen
  \bibfield  {author} {\bibinfo {author} {\bibfnamefont {M.}~\bibnamefont
  {Bartolozzi}}, \bibinfo {author} {\bibfnamefont {D.~B.}\ \bibnamefont
  {Leinweber}},\ and\ \bibinfo {author} {\bibfnamefont {A.~W.}\ \bibnamefont
  {Thomas}},\ }\href@noop {} {\bibfield  {journal} {\bibinfo  {journal} {Phys.
  Rev. E}\ }\textbf {\bibinfo {volume} {72}},\ \bibinfo {pages} {046113}
  (\bibinfo {year} {2005})}\BibitemShut {NoStop}%
\bibitem [{\citenamefont {Jacobmeier}(2005)}]{Jacobmeier05}%
  \BibitemOpen
  \bibfield  {author} {\bibinfo {author} {\bibfnamefont {D.}~\bibnamefont
  {Jacobmeier}},\ }\href@noop {} {\bibfield  {journal} {\bibinfo  {journal}
  {Int. J. Mod. Phys. C}\ }\textbf {\bibinfo {volume} {16}},\ \bibinfo {pages}
  {633} (\bibinfo {year} {2005})}\BibitemShut {NoStop}%
\bibitem [{\citenamefont {Dinkelberg}\ \emph {et~al.}(2021)\citenamefont
  {Dinkelberg}, \citenamefont {Mac{C}arron}, \citenamefont {Maher},\ and\
  \citenamefont {Quayle}}]{DMMQ21}%
  \BibitemOpen
  \bibfield  {author} {\bibinfo {author} {\bibfnamefont {A.}~\bibnamefont
  {Dinkelberg}}, \bibinfo {author} {\bibfnamefont {P.}~\bibnamefont
  {Mac{C}arron}}, \bibinfo {author} {\bibfnamefont {P.~J.}\ \bibnamefont
  {Maher}},\ and\ \bibinfo {author} {\bibfnamefont {M.}~\bibnamefont
  {Quayle}},\ }\href@noop {} {\bibfield  {journal} {\bibinfo  {journal}
  {Physica A}\ }\textbf {\bibinfo {volume} {578}},\ \bibinfo {pages} {126086}
  (\bibinfo {year} {2021})}\BibitemShut {NoStop}%
\bibitem [{\citenamefont {Lorenz}(2007)}]{Lorenz07}%
  \BibitemOpen
  \bibfield  {author} {\bibinfo {author} {\bibfnamefont {J.}~\bibnamefont
  {Lorenz}},\ }\href@noop {} {\bibfield  {journal} {\bibinfo  {journal} {Int.
  J. Mod. Phys. C}\ }\textbf {\bibinfo {volume} {18}},\ \bibinfo {pages} {1819}
  (\bibinfo {year} {2007})}\BibitemShut {NoStop}%
\bibitem [{\citenamefont {Bernardo}\ \emph {et~al.}(2024)\citenamefont
  {Bernardo}, \citenamefont {Altafini}, \citenamefont {Proskurnikov},\ and\
  \citenamefont {Vasca}}]{BCReview}%
  \BibitemOpen
  \bibfield  {author} {\bibinfo {author} {\bibfnamefont {C.}~\bibnamefont
  {Bernardo}}, \bibinfo {author} {\bibfnamefont {C.}~\bibnamefont {Altafini}},
  \bibinfo {author} {\bibfnamefont {A.}~\bibnamefont {Proskurnikov}},\ and\
  \bibinfo {author} {\bibfnamefont {F.}~\bibnamefont {Vasca}},\ }\href@noop {}
  {\bibfield  {journal} {\bibinfo  {journal} {Automatica}\ }\textbf {\bibinfo
  {volume} {159}},\ \bibinfo {pages} {111302} (\bibinfo {year}
  {2024})}\BibitemShut {NoStop}%
\bibitem [{\citenamefont {Fortunato}\ \emph {et~al.}(2005)\citenamefont
  {Fortunato}, \citenamefont {Latora}, \citenamefont {Pluchino},\ and\
  \citenamefont {Rapisarda}}]{FLPR05}%
  \BibitemOpen
  \bibfield  {author} {\bibinfo {author} {\bibfnamefont {S.}~\bibnamefont
  {Fortunato}}, \bibinfo {author} {\bibfnamefont {V.}~\bibnamefont {Latora}},
  \bibinfo {author} {\bibfnamefont {A.}~\bibnamefont {Pluchino}},\ and\
  \bibinfo {author} {\bibfnamefont {A.}~\bibnamefont {Rapisarda}},\ }\href@noop
  {} {\bibfield  {journal} {\bibinfo  {journal} {Int. J. Mod. Phys. C}\
  }\textbf {\bibinfo {volume} {16}},\ \bibinfo {pages} {1535} (\bibinfo {year}
  {2005})}\BibitemShut {NoStop}%
\bibitem [{\citenamefont {Lorenz}(2008)}]{Lorenz08}%
  \BibitemOpen
  \bibfield  {author} {\bibinfo {author} {\bibfnamefont {J.}~\bibnamefont
  {Lorenz}},\ }in\ \href@noop {} {\emph {\bibinfo {booktitle} {Managing
  Complexity: {I}nsights, Concepts, Applications}}},\ \bibinfo {editor} {edited
  by\ \bibinfo {editor} {\bibfnamefont {D.}~\bibnamefont {Helbing}}}\ (\bibinfo
   {publisher} {Springer},\ \bibinfo {address} {Berlin},\ \bibinfo {year}
  {2008})\ pp.\ \bibinfo {pages} {321--334}\BibitemShut {NoStop}%
\bibitem [{\citenamefont {Huet}\ and\ \citenamefont {Deffuant}(2010)}]{HD10}%
  \BibitemOpen
  \bibfield  {author} {\bibinfo {author} {\bibfnamefont {S.}~\bibnamefont
  {Huet}}\ and\ \bibinfo {author} {\bibfnamefont {G.}~\bibnamefont
  {Deffuant}},\ }\href@noop {} {\bibfield  {journal} {\bibinfo  {journal} {Adv.
  Complex Sys.}\ }\textbf {\bibinfo {volume} {13}},\ \bibinfo {pages} {405}
  (\bibinfo {year} {2010})}\BibitemShut {NoStop}%
\bibitem [{\citenamefont {Liu}\ \emph {et~al.}(2023)\citenamefont {Liu},
  \citenamefont {M{\"a}s}, \citenamefont {Xia},\ and\ \citenamefont
  {Flache}}]{LMXF23}%
  \BibitemOpen
  \bibfield  {author} {\bibinfo {author} {\bibfnamefont {S.}~\bibnamefont
  {Liu}}, \bibinfo {author} {\bibfnamefont {M.}~\bibnamefont {M{\"a}s}},
  \bibinfo {author} {\bibfnamefont {H.}~\bibnamefont {Xia}},\ and\ \bibinfo
  {author} {\bibfnamefont {A.}~\bibnamefont {Flache}},\ }\href@noop {}
  {\bibfield  {journal} {\bibinfo  {journal} {J. Artifical Societies and Social
  Simulation}\ }\textbf {\bibinfo {volume} {26}},\ \bibinfo {pages} {8}
  (\bibinfo {year} {2023})}\BibitemShut {NoStop}%
\bibitem [{\citenamefont {Zou}(2023)}]{Myproject}%
  \BibitemOpen
  \bibfield  {author} {\bibinfo {author} {\bibfnamefont {X.}~\bibnamefont
  {Zou}},\ }\href {https://www.jasss.org/5/3/2.html} {\emph {\bibinfo {title}
  {Opinion Dynamics: Go Beyond {A}xelrod Model and the Application on Voting
  Dynamics}}},\ \bibinfo {type} {Project report of the course ``Data Analysis
  And Modeling In Physics''}\ (\bibinfo  {institution} {Univ. of Hong Kong},\
  \bibinfo {year} {2023})\BibitemShut {NoStop}%
\bibitem [{\citenamefont {Barabási}(2009)}]{scalefreesociety1}%
  \BibitemOpen
  \bibfield  {author} {\bibinfo {author} {\bibfnamefont {A.-L.}\ \bibnamefont
  {Barabási}},\ }\href@noop {} {\bibfield  {journal} {\bibinfo  {journal}
  {Science}\ }\textbf {\bibinfo {volume} {325}},\ \bibinfo {pages} {412}
  (\bibinfo {year} {2009})}\BibitemShut {NoStop}%
\bibitem [{\citenamefont {Hovland}\ \emph {et~al.}(1953)\citenamefont
  {Hovland}, \citenamefont {Janis},\ and\ \citenamefont {Kelley}}]{HJK53}%
  \BibitemOpen
  \bibfield  {author} {\bibinfo {author} {\bibfnamefont {C.~I.}\ \bibnamefont
  {Hovland}}, \bibinfo {author} {\bibfnamefont {I.~L.}\ \bibnamefont {Janis}},\
  and\ \bibinfo {author} {\bibfnamefont {H.~H.}\ \bibnamefont {Kelley}},\
  }\href@noop {} {\emph {\bibinfo {title} {Communication And Persuasion;
  Psychological Studies Of Opinion Change}}}\ (\bibinfo  {publisher} {Yale U
  Press},\ \bibinfo {year} {1953})\BibitemShut {NoStop}%
\bibitem [{\citenamefont {Levy}\ and\ \citenamefont
  {Maaravi}(2017)}]{Boomerang_effect1}%
  \BibitemOpen
  \bibfield  {author} {\bibinfo {author} {\bibfnamefont {A.}~\bibnamefont
  {Levy}}\ and\ \bibinfo {author} {\bibfnamefont {Y.}~\bibnamefont {Maaravi}},\
  }\href@noop {} {\bibfield  {journal} {\bibinfo  {journal} {Soc. Influ.}\
  }\textbf {\bibinfo {volume} {13}},\ \bibinfo {pages} {39} (\bibinfo {year}
  {2017})}\BibitemShut {NoStop}%
\bibitem [{\citenamefont {Ringold}(2002)}]{Boomerang_effect2}%
  \BibitemOpen
  \bibfield  {author} {\bibinfo {author} {\bibfnamefont {D.}~\bibnamefont
  {Ringold}},\ }\href@noop {} {\bibfield  {journal} {\bibinfo  {journal} {J.
  Consumer Policy}\ }\textbf {\bibinfo {volume} {25}},\ \bibinfo {pages} {27}
  (\bibinfo {year} {2002})}\BibitemShut {NoStop}%
\bibitem [{\citenamefont {Ecker}\ \emph {et~al.}(2022)\citenamefont {Ecker},
  \citenamefont {Lewandowsky}, \citenamefont {Cook}, \citenamefont {Schmid},
  \citenamefont {Fazio}, \citenamefont {Brashier}, \citenamefont {Kendeou},
  \citenamefont {Vraga},\ and\ \citenamefont
  {Amazeen}}]{Boomerang_misinformation_increase}%
  \BibitemOpen
  \bibfield  {author} {\bibinfo {author} {\bibfnamefont {U.~K.~H.}\
  \bibnamefont {Ecker}}, \bibinfo {author} {\bibfnamefont {S.}~\bibnamefont
  {Lewandowsky}}, \bibinfo {author} {\bibfnamefont {J.}~\bibnamefont {Cook}},
  \bibinfo {author} {\bibfnamefont {P.}~\bibnamefont {Schmid}}, \bibinfo
  {author} {\bibfnamefont {L.~K.}\ \bibnamefont {Fazio}}, \bibinfo {author}
  {\bibfnamefont {N.}~\bibnamefont {Brashier}}, \bibinfo {author}
  {\bibfnamefont {P.}~\bibnamefont {Kendeou}}, \bibinfo {author} {\bibfnamefont
  {E.~K.}\ \bibnamefont {Vraga}},\ and\ \bibinfo {author} {\bibfnamefont
  {M.~A.}\ \bibnamefont {Amazeen}},\ }\href
  {https://doi.org/10.1038/s44159-021-00006-y} {\bibfield  {journal} {\bibinfo
  {journal} {Nat. Rev. Psychol.}\ }\textbf {\bibinfo {volume} {1}},\ \bibinfo
  {pages} {13} (\bibinfo {year} {2022})}\BibitemShut {NoStop}%
\bibitem [{\citenamefont {Schultz}\ \emph {et~al.}(2007)\citenamefont
  {Schultz}, \citenamefont {Nolan}, \citenamefont {Cialdini}, \citenamefont
  {Goldstein},\ and\ \citenamefont {Griskevicius}}]{SNCGG07}%
  \BibitemOpen
  \bibfield  {author} {\bibinfo {author} {\bibfnamefont {P.~W.}\ \bibnamefont
  {Schultz}}, \bibinfo {author} {\bibfnamefont {J.~M.}\ \bibnamefont {Nolan}},
  \bibinfo {author} {\bibfnamefont {R.~B.}\ \bibnamefont {Cialdini}}, \bibinfo
  {author} {\bibfnamefont {N.~J.}\ \bibnamefont {Goldstein}},\ and\ \bibinfo
  {author} {\bibfnamefont {V.}~\bibnamefont {Griskevicius}},\ }\href@noop {}
  {\bibfield  {journal} {\bibinfo  {journal} {Psychol. Sci.}\ }\textbf
  {\bibinfo {volume} {18}},\ \bibinfo {pages} {429} (\bibinfo {year}
  {2007})}\BibitemShut {NoStop}%
\bibitem [{\citenamefont {Verkooijen}\ \emph {et~al.}(2015)\citenamefont
  {Verkooijen}, \citenamefont {Stok},\ and\ \citenamefont {Moolen}}]{TSM15}%
  \BibitemOpen
  \bibfield  {author} {\bibinfo {author} {\bibfnamefont {K.~T.}\ \bibnamefont
  {Verkooijen}}, \bibinfo {author} {\bibfnamefont {F.~M.}\ \bibnamefont
  {Stok}},\ and\ \bibinfo {author} {\bibfnamefont {S.}~\bibnamefont {Moolen}},\
  }\href@noop {} {\bibfield  {journal} {\bibinfo  {journal} {Eur. J. Soc.
  Psychol.}\ }\textbf {\bibinfo {volume} {45}},\ \bibinfo {pages} {417}
  (\bibinfo {year} {2015})}\BibitemShut {NoStop}%
\bibitem [{\citenamefont {Liu}\ \emph {et~al.}(2020)\citenamefont {Liu},
  \citenamefont {Liu}, \citenamefont {Wang},\ and\ \citenamefont
  {Xu}}]{LLWX19}%
  \BibitemOpen
  \bibfield  {author} {\bibinfo {author} {\bibfnamefont {A.~Z.}\ \bibnamefont
  {Liu}}, \bibinfo {author} {\bibfnamefont {A.~X.}\ \bibnamefont {Liu}},
  \bibinfo {author} {\bibfnamefont {R.}~\bibnamefont {Wang}},\ and\ \bibinfo
  {author} {\bibfnamefont {S.~X.}\ \bibnamefont {Xu}},\ }\href@noop {}
  {\bibfield  {journal} {\bibinfo  {journal} {J. Manag. Stud.}\ }\textbf
  {\bibinfo {volume} {57}},\ \bibinfo {pages} {1437} (\bibinfo {year}
  {2020})}\BibitemShut {NoStop}%
\bibitem [{\citenamefont {Tak{\'a}cs}\ \emph {et~al.}(2016)\citenamefont
  {Tak{\'a}cs}, \citenamefont {Flache},\ and\ \citenamefont {M{\"a}s}}]{TFM16}%
  \BibitemOpen
  \bibfield  {author} {\bibinfo {author} {\bibfnamefont {K.}~\bibnamefont
  {Tak{\'a}cs}}, \bibinfo {author} {\bibfnamefont {A.}~\bibnamefont {Flache}},\
  and\ \bibinfo {author} {\bibfnamefont {M.}~\bibnamefont {M{\"a}s}},\
  }\href@noop {} {\bibfield  {journal} {\bibinfo  {journal} {PLOS One}\
  }\textbf {\bibinfo {volume} {11}},\ \bibinfo {pages} {e0157948} (\bibinfo
  {year} {2016})}\BibitemShut {NoStop}%
\bibitem [{\citenamefont {Bail}\ \emph {et~al.}(2018)\citenamefont {Bail},
  \citenamefont {Argyle}, \citenamefont {Brown}, \citenamefont {Bumpus},
  \citenamefont {Chen}, \citenamefont {Hunzaker}, \citenamefont {Lee},
  \citenamefont {Mann}, \citenamefont {Merhout},\ and\ \citenamefont
  {Volfovsky}}]{Bail18}%
  \BibitemOpen
  \bibfield  {author} {\bibinfo {author} {\bibfnamefont {C.~A.}\ \bibnamefont
  {Bail}}, \bibinfo {author} {\bibfnamefont {L.~P.}\ \bibnamefont {Argyle}},
  \bibinfo {author} {\bibfnamefont {T.~W.}\ \bibnamefont {Brown}}, \bibinfo
  {author} {\bibfnamefont {J.~P.}\ \bibnamefont {Bumpus}}, \bibinfo {author}
  {\bibfnamefont {H.}~\bibnamefont {Chen}}, \bibinfo {author} {\bibfnamefont
  {M.~B.~F.}\ \bibnamefont {Hunzaker}}, \bibinfo {author} {\bibfnamefont
  {J.}~\bibnamefont {Lee}}, \bibinfo {author} {\bibfnamefont {M.}~\bibnamefont
  {Mann}}, \bibinfo {author} {\bibfnamefont {F.}~\bibnamefont {Merhout}},\ and\
  \bibinfo {author} {\bibfnamefont {A.}~\bibnamefont {Volfovsky}},\ }\href@noop
  {} {\bibfield  {journal} {\bibinfo  {journal} {Proc. Natl. Acad. Sci.
  U.S.A.}\ }\textbf {\bibinfo {volume} {115}},\ \bibinfo {pages} {9216}
  (\bibinfo {year} {2018})}\BibitemShut {NoStop}%
\bibitem [{\citenamefont {Pan}\ \emph {et~al.}(2025)\citenamefont {Pan},
  \citenamefont {Blut}, \citenamefont {Ghiassaleh},\ and\ \citenamefont
  {Lee}}]{PBGL25}%
  \BibitemOpen
  \bibfield  {author} {\bibinfo {author} {\bibfnamefont {M.}~\bibnamefont
  {Pan}}, \bibinfo {author} {\bibfnamefont {M.}~\bibnamefont {Blut}}, \bibinfo
  {author} {\bibfnamefont {A.}~\bibnamefont {Ghiassaleh}},\ and\ \bibinfo
  {author} {\bibfnamefont {Z.~W.~Y.}\ \bibnamefont {Lee}},\ }\href@noop {}
  {\bibfield  {journal} {\bibinfo  {journal} {J. Acad. Mark. Sci.}\ }\textbf
  {\bibinfo {volume} {53}},\ \bibinfo {pages} {52} (\bibinfo {year}
  {2025})}\BibitemShut {NoStop}%
\bibitem [{\citenamefont {Conde}\ and\ \citenamefont {Casais}(2023)}]{CC23}%
  \BibitemOpen
  \bibfield  {author} {\bibinfo {author} {\bibfnamefont {R.}~\bibnamefont
  {Conde}}\ and\ \bibinfo {author} {\bibfnamefont {B.}~\bibnamefont {Casais}},\
  }\href@noop {} {\bibfield  {journal} {\bibinfo  {journal} {J. Bus. Res.}\
  }\textbf {\bibinfo {volume} {158}},\ \bibinfo {pages} {113708} (\bibinfo
  {year} {2023})}\BibitemShut {NoStop}%
\bibitem [{\citenamefont {Peter}\ and\ \citenamefont {Muth}(2023)}]{PM23}%
  \BibitemOpen
  \bibfield  {author} {\bibinfo {author} {\bibfnamefont {C.}~\bibnamefont
  {Peter}}\ and\ \bibinfo {author} {\bibfnamefont {L.}~\bibnamefont {Muth}},\
  }\href@noop {} {\bibfield  {journal} {\bibinfo  {journal} {Media Commun.}\
  }\textbf {\bibinfo {volume} {11}},\ \bibinfo {pages} {164} (\bibinfo {year}
  {2023})}\BibitemShut {NoStop}%
\bibitem [{\citenamefont {Liu}\ and\ \citenamefont {Zheng}(2024)}]{LZ24}%
  \BibitemOpen
  \bibfield  {author} {\bibinfo {author} {\bibfnamefont {X.}~\bibnamefont
  {Liu}}\ and\ \bibinfo {author} {\bibfnamefont {X.}~\bibnamefont {Zheng}},\
  }\href@noop {} {\bibfield  {journal} {\bibinfo  {journal} {Humanit. Soc.
  Sci.}\ }\textbf {\bibinfo {volume} {11}},\ \bibinfo {pages} {15} (\bibinfo
  {year} {2024})}\BibitemShut {NoStop}%
\bibitem [{\citenamefont {Zou}(2025)}]{pythoncode}%
  \BibitemOpen
  \bibfield  {author} {\bibinfo {author} {\bibfnamefont {X.}~\bibnamefont
  {Zou}},\ }\href
  {https://github.com/SimbaZouXiang/Sociophysics_Option_dynamics} {\bibinfo
  {title} {Python code used in our simulation}},\ \bibinfo {howpublished}
  {available in
  \url{https://github.com/SimbaZouXiang/Sociophysics_Option_dynamics}}
  (\bibinfo {year} {2025})\BibitemShut {NoStop}%
\end{thebibliography}%

\end{document}